\documentclass[fleqn,usenatbib]{mnras}

\usepackage{newtxtext,newtxmath}
\usepackage[T1]{fontenc}
\usepackage{ae,aecompl}
\usepackage[usenames,dvipsnames,svgnames,hyperref]{xcolor}
\usepackage{soul,ulem}

\usepackage{graphicx}	% Including figure files
\usepackage{amsmath}	% Advanced maths commands
\usepackage{subfigure}
\usepackage{multirow}
\usepackage{enumitem}
\usepackage{xfrac}
\usepackage{tikz}
\usetikzlibrary{arrows.meta}
\tikzset{>={Latex[width = 1mm,length = 2mm]},
    base/.style = {rectangle, draw = RoyalBlue, thick, minimum width = 8cm, minimum height = 1cm, text centered, font = \sffamily}, 
    repeat1/.style = {base, rounded corners, draw = RedOrange, minimum width = 7cm}, 
    repeat2/.style = {repeat1, draw = YellowGreen, minimum width = 6cm}
}

\def\code#1{{\tt \textsc{#1}}}
\def\clustar{\code{CluSTAR-ND}}
\def\astrolink{\code{AstroLink}}
\def\hoptics{\code{Halo-OPTICS}}
\def\optics{\code{OPTICS}}
\usepackage{stfloats}
\title[Hierarchical Clusters in Haloes]{The Hierarchical Structure of Galactic Haloes: 
\\
Differentiating Clusters from Stochastic Clumping with \textsc{AstroLink}}
\author[Oliver et al.]{
William H. Oliver$^{1, 2, 3}$\thanks{E-mail: william.oliver@iwr.uni-heidelberg.de},
Pascal J. Elahi$^{4}$,
Geraint F. Lewis$^{3}$, and
Tobias Buck$^{1, 2}$
\\
$^{1}$Universit\"{a}t Heidelberg, Interdisziplin\"{a}res Zentrum f\"{u}r Wissenschaftliches Rechnen, Im Neuenheimer Feld 205, D-69120 Heidelberg, Germany\\
$^{2}$Universit\"{a}t Heidelberg, Zentrum f\"{u}r Astronomie, Institut f\"{u}r Theoretische Astrophysik, Albert-Ueberle-Stra{\ss}e 2, D-69120 Heidelberg, Germany\\
$^{3}$Sydney Institute for Astronomy, School of Physics A28, The University of Sydney, NSW 2006, Australia\\
$^{4}$Pawsey Supercomputing Research Centre, 1 Bryce Avenue, Kensington, WA 6151, Australia
}

\date{Accepted XXX. Received YYY; in original form ZZZ}

\pubyear{2023}

\begin{document}
\label{firstpage}
\pagerange{\pageref{firstpage}--\pageref{lastpage}}
\maketitle

\begin{abstract}
We present \astrolink, an efficient and versatile clustering algorithm designed to hierarchically classify astrophysically-relevant structures from both synthetic and observational data sets. We build upon \clustar, a hierarchical galaxy/(sub)halo finder, so that \astrolink\ now generates a two-dimensional representation of the implicit clustering structure as well as ensuring that clusters are statistically distinct from the noisy density fluctuations implicit within the $n$-dimensional input data. This redesign replaces the three cluster extraction parameters from \clustar\ with a single parameter, $S$ -- the lower statistical significance threshold of clusters, which can be automatically and reliably estimated via a dynamical model-fitting process. We demonstrate the robustness of this approach compared to \astrolink's predecessors by applying each algorithm to a suite of simulated galaxies defined over various feature spaces. We find that \astrolink\ delivers a more powerful clustering performance while being $\sim27\%$ faster and using less memory than \clustar. With these improvements, \astrolink\ is ideally suited to extracting a meaningful set of hierarchical and arbitrarily-shaped astrophysical clusters from both synthetic and observational data sets -- lending itself as a great tool for morphological decomposition within the context of hierarchical structure formation.
\end{abstract}


%%%%%%%%%%%%%%%%%%%%%%%%%%%%%%%%%%%%%%%%%%%%%%%%%%

%%%%%%%%%%%%%%%%% BODY OF PAPER %%%%%%%%%%%%%%%%%%

\section{Introduction} \label{sec:introduction}
%Correctly classifying astrophysically-relevant structures in observational data sets is crucial for advancing our understanding of the Universe. Identifying spatial, kinematic, and chemical overdensities within these data sets allows us to make predictions about structure formation and evolution. With an ever-increasing volume of data, it is essential to adopt data-mining approaches for astrophysical structure identification, as manual inspection methods cannot reliably be used to identify all relevant structure.
Correctly identifying spatial, kinematic, and chemical structures within simulated and observational data sets is crucial for advancing our understanding of the Universe and allows us to make predictions about structure formation and evolution. With an ever-increasing volume of data, it is essential to adopt data-mining approaches for astrophysical structure identification, as manual inspection methods cannot reliably be used to identify all relevant structure.

%Historically, structures were discovered through direct data inspection, such as galaxies and galactic clusters with photographic plates \citep{Abell1958}, or more recently, galactic substructure has been uncovered using specific data projections \citep[e.g.][]{Arifyanto2006, Duffau2006, Williams2011, Helmi2017, Belokurov2018}. However, w

Data mining algorithms that find structure are referred to as clustering algorithms. Clustering algorithms, used for identifying astrophysical structures in observational data, often incorporate physical models to classify groups. These models commonly include constraints on orbital motion influenced by the parent structure's gravitational potential, as seen in algorithms such as \code{StreamFinder} \citep{Malhan2018a} and the \code{xGC3} suite \citep{Johnston1996, Mateu2011, Mateu2017}. They may also apply data projections or transformations to target specific structure types, e.g. \code{StarGO} \citep{Yuan2018}, \code{HSS} \citep{Pearson2022}, and \code{Via Machinae} \citep{Shih2022}. While these algorithms effectively reveal the structures they are designed to target, these constraints limit their ability to discover a broad range of structure types and their inter-relationships.

Clustering algorithms built for application to synthetic data -- often called galaxy/(sub)halo finders -- can be categorized into three types: configuration space finders e.g. \code{SubFind} \citep{Springel2001}, \code{AHF} \citep{Knollmann2009}, and \code{CompaSO} \citep{Hadzhiyska2021}, phase space finders e.g. \code{6DFOF} \citep{Diemand2006}, \code{HSF} \citep{Maciejewski2009}, \code{ROCKSTAR} \citep{Behroozi2012}, and \code{VELOCIraptor} \citep{Elahi2019}, and tracking finders e.g. \code{SURV} \citep{Tormen2004, Giocoli2008} and \code{HBT+} \citep{Han2018}. Configuration space finders primarily use 3D spatial particle positions to identify physical overdensities, particle velocities are also often incorporated to refine these self-bound haloes. Phase space finders use both 3D spatial positions and 3D velocities of particles to find clusters. While tracking finders use either configuration or phase space to construct haloes but are assisted by particle tracking in their determination of such groups at later times.

While simulation-specific astrophysical clustering algorithms have evolved over recent decades, many still rely on the Spherical Overdensity \citep[\code{SO};][]{Press1974} or the Friends-Of-Friends \citep[\code{FOF};][]{Davis1985} algorithms at their core. The \code{SO} algorithm constructs galaxies and (sub)haloes by locating spatial density peaks and expanding spherical volumes from them until a specified overdensity is reached. The \code{FOF} algorithm groups particles connected by distances less than the linking length ($l_x$), often set at $0.2l_\mathrm{mean}$ (corresponding to overdensities with $\Delta \approx 200$), where $l_\mathrm{mean}$ is the mean particle separation within the simulation box. These algorithms have established themselves as the primary means of defining galaxies and (sub)haloes, especially within synthetic data. This isn't inherently negative, as structures identified by these methods, followed by an unbinding process, represent true self-bound groups. Moreover, numerous comparative studies have shown strong agreement among modern galaxy/(sub)halo finders, including those built on the \code{SO} and \code{FOF} algorithms \citep{Knebe2011, Knebe2013a, Onions2012, Onions2013, Elahi2013, Avila2014, Lee2014, Behroozi2015}. However, the \code{SO} and \code{FOF} algorithms may overlook diffuse debris or anisotropic clusters and do not produce hierarchical clusterings. This often necessitates their iterative application in order to identify clusters from structure with diverse characteristics. Consequently, modern galaxy/(sub)halo finders may struggle to capture all relevant structures, though some are better able to identify more that just self-bound objects \cite{Elahi2013}.

A versatile clustering algorithm is needed to overcome the limitations of both observation-specific and simulation-specific astrophysical clustering methods while preserving their advantages. Such an algorithm should be applicable to both observational and synthetic data sets of varying sizes and feature types, without imposing model constraints. It should also provide an adaptive measure of hierarchical clustering structure, accommodating the discovery of structures at different overdensity levels. Generalised algorithms such as \code{OPTICS} \citep{Ankerst1999} and \code{HDBSCAN} \citep{Campello2015, McInnes2017} have recently gained popularity in astrophysical structure discovery -- e.g. \citet{Costado2016, McConnachie2018, Canovas2019, Massaro2019, Ward2020, Higgs2021, Jensen2021, Soto2022} for \code{OPTICS} and \citet{Ruiz2018, Mahajan2018, Koppelman2019, Jayasinghe2019, Kounkel2019, Webb2020, Kamdar2021, Walmsley2022, Lovdal2022, Casamiquela2022} for \code{HDBSCAN}. Some astrophysical codes, e.g. \code{HOP} \citep{Eisenstein1998a} and \code{ADAPTAHOP} \citep{Aubert2004}, share similarities to these algorithms -- however, neither \code{OPTICS} nor \code{HDBSCAN} are designed with astrophysics specifically in mind -- in fact, few generalised codes have been e.g. \code{EnLink} \citep{Sharma2009}, \code{FOPTICS} \citep{Fuentes2017}, and \clustar\ \citep{Oliver2022}.
%FOPTICS footnote \footnote{Although this algorithm was designed to be used on the position-velocity phase-space of stars, rather than some arbitrary combination of informative features.}

We develop a novel and almost entirely data driven astrophysical clustering algorithm \astrolink\ by improving upon the \clustar\ algorithm. First, we briefly summarise the \clustar\ algorithm in Sec. \ref{sec:clustar}. We then describe the \astrolink\ algorithm throughout Sec. \ref{sec:clustarr}. In Sec. \ref{sec:performance} we show the algorithm in practice and compare it to its predecessors, \clustar\ and \hoptics\ \citep{Oliver2020}. Then finally, we make our conclusions and present our ideas for future work in Sec. \ref{sec:conclusion}.

\section{An Overview of the CluSTAR-ND Algorithm} \label{sec:clustar}
The \textbf{Clu}stering \textbf{S}tructure via \textbf{T}ransformative \textbf{A}ggregation and \textbf{R}ejection in \textbf{N}-\textbf{D}imensions algorithm \citep[\clustar; ][]{Oliver2022} generates a hierarchical clustering that represents galaxies and their substructures from input data of any size and number of features. This algorithm improves on \hoptics\ \citep{Oliver2020} by enhancing its computational efficiency and extending its applicability to $n$-dimensional data sets. \hoptics\ suitably adapts the \optics\ algorithm \citep{Ankerst1999} for use with astrophysical data sets. Furthermore, \optics\ is itself a hierarchical extension of the \code{DBSCAN} algorithm \citep{Ester1996}, which is conceptually an extension of the \code{FOF} algorithm that substitutes the point-point linking scheme for the more robust neighbourhood-neighbourhood linking scheme. Due to this chain of algorithm succession, \clustar\ exhibits elements from each of these algorithms in the following approach that:
\begin{enumerate}[leftmargin = *, align = left]
    \item \label{step:i} Optionally identifies FOF field haloes from the input data (otherwise treating the input data as a single FOF halo);
    \item \label{step:ii} Optionally transforms the $n$-dimensional field halo data via a Principle Component Analysis (PCA) transformation;
    \item \label{step:iii} Estimates the local density field from the transformed data;
    \item \label{step:iv} Initiates groups with points at local maxima within this field;
    \item \label{step:v} Aggregates points in order of decreasing density to the group whose members belong to that point's neighbourhood;
    \item \label{step:vi} Merges groups when at least one of each of their members are part of a next-to-be-aggregated point's neighbourhood;
    \item \label{step:vii} Concurrently monitors whether a group, just prior to merging, meets \textit{cluster} criteria;
    \item \label{step:viii} Removes outliers from clusters;
    \item \label{step:ix} And then optionally repeats steps \ref{step:ii} -- \ref{step:ix} using the first level of the hierarchy as input data.
\end{enumerate}
Step \ref{step:i} allows users to ensure that root-level clusters in the hierarchy are galaxies/field haloes. While step \ref{step:ii} optionally removes any unwanted global (intra-galaxy/halo) over-dependencies on a subset of features before searching for substructure -- i.e. if {\tt adaptive} $\geq 1$, a PCA transformation is applied and the components are re-scaled with unit variance per component. Step \ref{step:iii} uses each point's $k_\mathrm{den}$ nearest neighbours to estimate the local density -- reducing the complexity of an $n$-dimensional data set to the simplicity of a scalar value. The set of each point's $k_\mathrm{link}$ ($\leq k_\mathrm{den}$) nearest neighbours are then used in steps \ref{step:iv} -- \ref{step:vi} to perform neighbourhood linkage.

In effect, steps \ref{step:iii} -- \ref{step:vi} enable \clustar\ to construct an analogue of the \optics\ reachability plot\footnote{\optics\ alone only provides a $2$-dimensional expression of the clustering structure within the input data -- this is called the reachability plot and is a plot of the \textit{reachability distance} vs the ordered index. Within this plot, clusters appear as valleys since they are more dense than their surrounds (smaller distances between points) and are ordered consecutively (local to each other). For more details on the \optics\ algorithm, refer to the original paper \citep{Ankerst1999} and \citet{Oliver2020}.}. Step \ref{step:vii} then ensures that the clustering hierarchy closely resembles subhaloes within galaxies. The clustering hierarchy of \clustar\ (and of \hoptics) is designed to reflect the cluster extraction processes of \citet{Sander2003}, \citet{Zhang2013}, and \citet{McConnachie2018} that act on the \optics\ reachability plot. As such;
\begin{enumerate}[leftmargin = *, align = left]
    \item[(a)] All clusters must have at least $k_\mathrm{link}$ points;
    \item[(b)] All clusters must have median densities that are at least $\rho_\mathrm{threshold}$ times that at their boundaries;
    \item[(c)] The hierarchy must not contain any lone leaf clusters;
    \item[(d)] And any parent-child pair of clusters must not share more than $f_\mathrm{reject}$ of the parent's points.
\end{enumerate}
\clustar\ automatically optimizes and selects the neighbourhood linkage size, $k_\mathrm{link}$, and the overdensity factor, $\rho_\mathrm{threshold}$, based on the user's choice of $k_\mathrm{den}$ as well as the dimensionality of the input data. The parameter $f_\mathrm{reject}$ is set at $0.9$ following the analysis performed in Sec. 3.3 in \citet{Oliver2020}.

After constructing the hierarchy, step \ref{step:viii} eliminates outliers from each cluster, ensuring that all remaining points have a density greater than $\rho_\mathrm{cut} \coloneq \mathrm{min}\{\rho_i \mid \mathrm{lof}(\rho_i) < S_\mathrm{outlier}\}$. Here $\mathrm{lof}(\rho_i)$ is defined in Eq. 8 of \citet{Oliver2022} and serves as a kernel density analogue of the local-outlier-factor formalised in \citet{Breunig1999}. The parameter $S_\mathrm{outlier}$ is optimally set to $2.5$, allowing for a moderate level of outlier removal without compromising the algorithm's clustering capability. Finally, if {\tt adaptive} $= 2$, step \ref{step:ix} can be used to adjust the governing distance metric for each level of substructure in the hierarchy so that clusters are dense relative to their parent cluster. For more details on \clustar, please refer to \citet{Oliver2022}.

\section{AstroLink: Building upon CluSTAR-ND} \label{sec:clustarr}
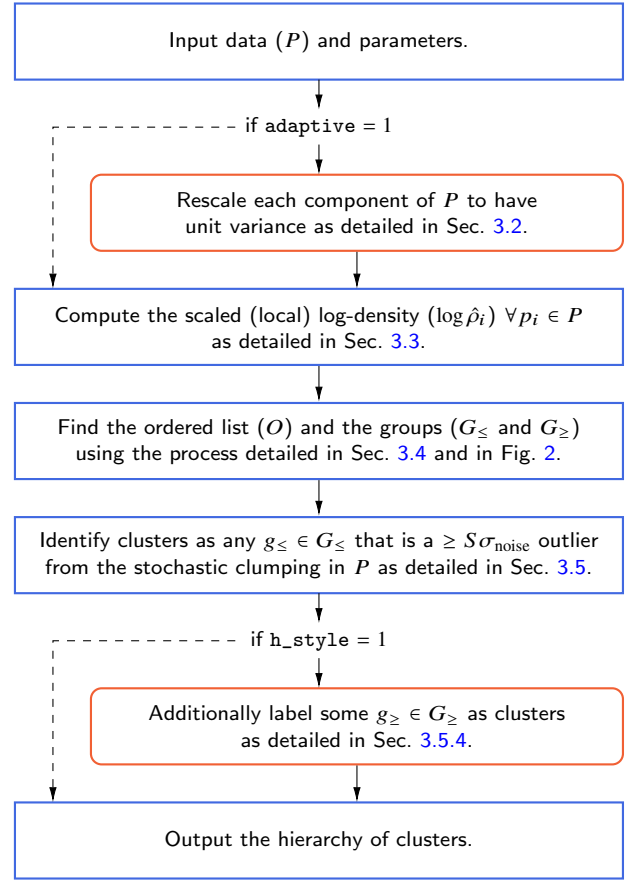
\begin{figure}
\begin{center}
\begin{tikzpicture}[node distance=1.5cm, every node/.style={fill=white, text centered, font=\sffamily}], align=center]
    % Specification of nodes (position, etc.)
    \node[align = center] (input)[base]
    {Input data ($P$) and parameters.};
    
    \node(ifpca)[text width = 2cm, below of=input, yshift=0.375cm]
    {if {\tt adaptive} $= 1$};
    
    \node[align = center] (pca)[repeat1, below of=ifpca, xshift=0.5cm, yshift=0.375cm]
    {Rescale each component of $P$ to have\\unit variance as detailed in Sec. \ref{subsec:transformation}.};
    
    \node[align = center] (density)[base, below of=pca, xshift=-0.5cm]
    {Compute the scaled (local) log-density ($\log\hat{\rho}_i$) $\forall p_i \in P$\\as detailed in Sec. \ref{subsec:density_estimation}.};
    
    \node[align = center] (aggregate)[base, below of=density]
    {Find the ordered list ($O$) and the groups ($G_\leq$ and $G_\geq$)\\using the process detailed in Sec. \ref{subsec:aggregation} and in Fig. \ref{fig:aggregation_visual}.};
    
    \node[align = center] (clusters)[base, below of=aggregate]
    {Identify clusters as any $g_\leq \in G_\leq$ that is a $\geq S\sigma_\mathrm{noise}$ outlier\\from the stochastic clumping in $P$ as detailed in Sec. \ref{subsec:identifyingclusters}.};

    \node(ifcorrection)[text width = 2cm, below of=clusters, yshift=0.375cm]
    {if {\tt h\_style} $= 1$};
    
    \node[align = center] (correction)[repeat1, below of=ifcorrection, xshift=0.5cm, yshift=0.375cm]
    {Additionally label some $g_\geq \in G_\geq$ as clusters\\as detailed in Sec. \ref{subsubsec:correction}.};
    
    \node[align = center] (output)[base, below of=correction, xshift=-0.5cm]
    {Output the hierarchy of clusters.};
    
    % Specification of lines between nodes specified above
    % with additional nodes for description
    \draw[<-](ifpca) -- +(0, 0.625);
    \draw[->](ifpca) -- +(0, -0.625);
    \draw[->](pca) -- +(0, -1);
    \draw[dashed, ->](ifpca) -- +(-3.5, 0) -- +(-3.5, -2.125);
    \draw[->](density) -- (aggregate);
    \draw[->](aggregate) -- (clusters);
    \draw[<-](ifcorrection) -- +(0, 0.625);
    \draw[->](ifcorrection) -- +(0, -0.625);
    \draw[->](correction) -- +(0, -1);
    \draw[dashed, ->](ifcorrection) -- +(-3.5, 0) -- +(-3.5, -2.125);
\end{tikzpicture}
\end{center}
\vspace{-8pt}
\caption{A high-level activity chart for the \astrolink\ algorithm. The nodes in the middle of the chart each correspond to specific subsections within Sec.~\ref{sec:clustarr} -- as indicated in the text of each node.}
\label{fig:searchsubstructure}
\end{figure}

\clustar\ performs well in many scenarios, but its cluster extraction process lacks the adaptability required to produce meaningful results from data sets with high or fluctuating noise levels. In these cases, users would be forced to manually adjust \clustar\ parameters, which is non-trivial due to the complex relationship between these parameters and the output. As such, we design \astrolink\ to remedy these drawbacks so that it can extract clusters with a single data-driven parameter, $S$ -- the lower threshold of statistical significance of clusters. This approach gives \astrolink\ the adaptability to accommodate different noise levels in the data while offering users an intuitive means of controlling the output. To maintain algorithmic transparency and for completeness, we now present each step of the \astrolink\ algorithm as shown in Fig. \ref{fig:searchsubstructure}, some of which are the same as is presented in \cite{Oliver2022} regarding \clustar.

\subsection{Root-level Clusters} \label{subsec:root_haloes}
In \clustar, the root-level clusters are found first using the \code{FOF} algorithm over the spatial positions -- ensuring that root-level clusters of the output are physically defined as galaxies/haloes. We have eliminated this step from \astrolink\ so that it can be seamlessly applied to various types of astrophysical data, including those that do not necessitate halo identification based on spatial coordinates before identifying substructure. Consequently, the output of \astrolink\ can be interpreted as a natural and mathematical clustering of the input data. If the user still requires the identification of galaxies/haloes prior to finding substructure, they should first apply an implementation of the \code{FOF} algorithm before then providing these results to \astrolink. Ordinarily, the input data is labelled as the root-level cluster -- making the hierarchy labelling of \astrolink\ compatible with \clustar.

\subsection{Transforming the Input Data} \label{subsec:transformation}
The first step of \astrolink\ is optional such that if {\tt adaptive} $= 1$ (default), the input data ($P$) is re-scaled so that each component has unit variance. This is used to remove any unwanted global over-dependencies (due to unit choice or otherwise) on a subset of $P$'s features before a search for substructure is conducted. This functions the same as the PCA transformation in step \ref{step:ii} of \clustar, but is faster as no rotation is being made (the output is unaffected by this since the Euclidean distance is rotation-invariant). Hence by calculating Euclidean distances on the transformed data, \astrolink\ effectively calculates Mahalanobis distances \citep{Mahalanobis1936} on the input data -- which guarantees that any resultant substructures will be dense with respect to the global shape of the input data -- although this may be sub-optimal if the input data is strongly multi-modal in a subset of the input data's feature space. To skip this step, the user can set {\tt adaptive} $= 0$, which may improve clustering results if the relative-scaling between dimensions is meaningful. Unlike in \clustar, \astrolink\ does not provide the option to set {\tt adaptive} $= 2$ and use a recursive PCA transformation (i.e. step \ref{step:ix}) as this was not shown to consistently increase clustering power.

\subsection{Estimating Local Density} \label{subsec:density_estimation}
After the optional transformation step, \astrolink\ estimates the local density of each point in a similar way to \clustar. First the set of $k_\mathrm{den}$ nearest neighbours, $N_{k_\mathrm{den}}$, are found for each data point -- the default value of which is $k_\mathrm{den} = 20$. The local density is then estimated by using these sets in conjunction with a multivariate Epanechnikov kernel\footnote{The Epanechnikov kernel is theoretically (asymptotically) optimal in that it minimizes the mean integrated squared error of the resulting density estimate.} \citep{Epanechnikov1969} and a balloon estimator \citep{Sain2002} such that;
\begin{align} \label{eq:densityestimator}
    &\rho_i \propto \frac{1}{h_i^d} \sum_{j \in N_{k_\mathrm{den}}} K\left(\frac{s(x_i, x_j)}{h_i}\right)\ \mathrm{where} \notag\\
    &h_i = \mathrm{max}\{s(x_i, x_j) \mid j \in N_{k_\mathrm{den}}\},\ \mathrm{and} \notag\\
    &K(u) \propto (1 - u^2).
\end{align}
Here $d$ is the dimensionality of the feature space, and $s(x_i, x_j)$ is the Euclidean distance between points $i$ and $j$ which, if {\tt adaptive} $= 1$, corresponds to the Mahalanobis distance on the input data. Note that $K(u) \coloneq 0$ for $u > 1$.

\begin{figure}
    \centering
    \includegraphics[trim={18.5mm 32mm 15mm 36.5mm}, clip, width=\columnwidth]{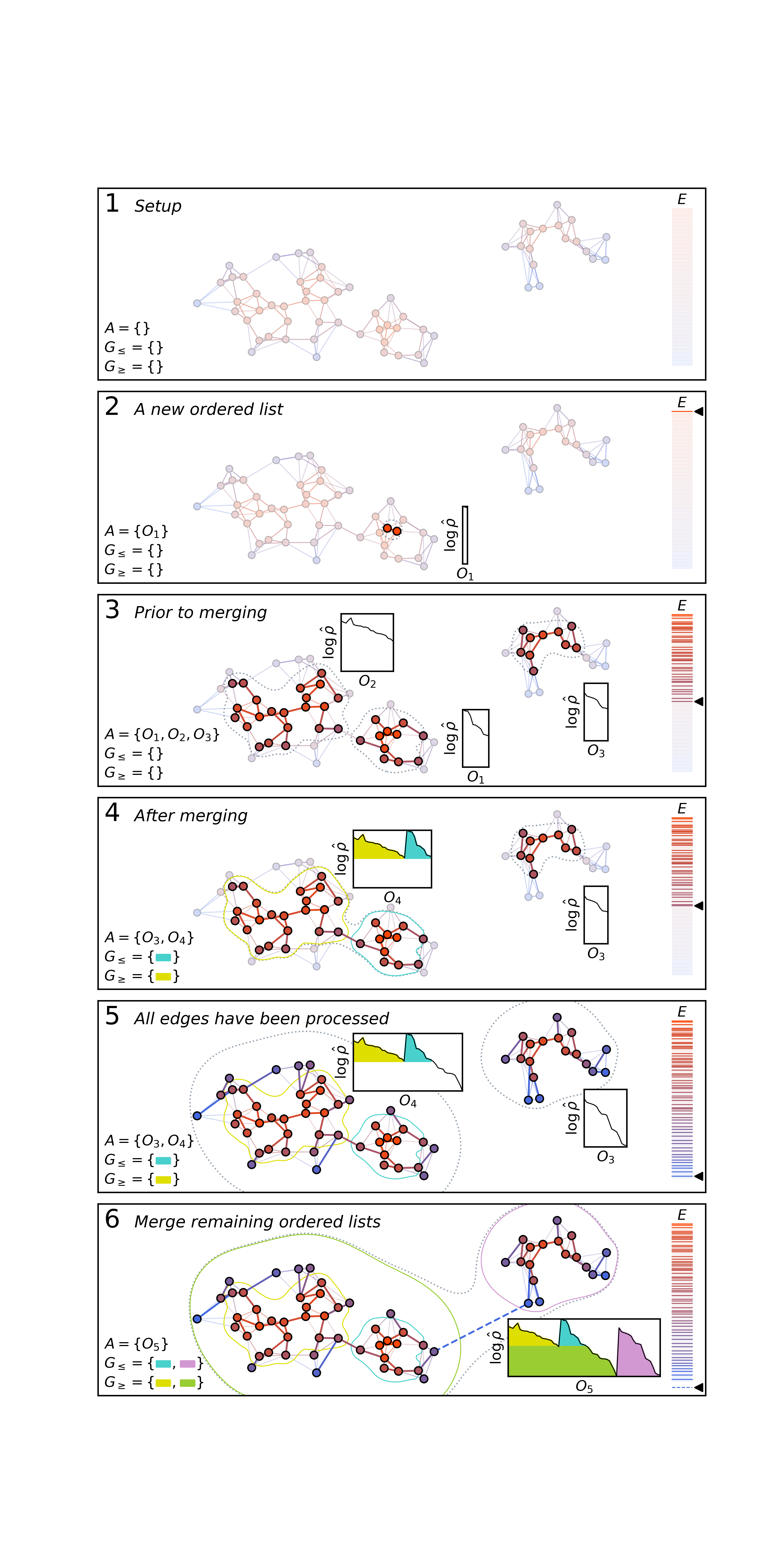}
    \vspace{-0.5cm}
    \caption{An illustration of the \astrolink\ aggregation process (using $k_\mathrm{den} = 10$ \& $k_\mathrm{link} = 5$) with new stages depicted in each panel. Here; a red-blue gradient indicates high-low $\log\hat\rho$ values/edge weightings; unused points/edges are transparent; dotted contour lines represent the growing iso-density surfaces (as in the analogy from Sec. \ref{subsec:aggregation}); and similarly, solid coloured lines represent those that envelope groups stored in $G_\leq$ and $G_\geq$ (these groups are also shown in the growing ordered-density plots within the inset panels). {\bf In panel 1}, the edges ($E$) have been computed and the empty sets ($A$, $G_\leq$, $G_\geq$) have been initialised. {\bf In panel 2}, the first edge has been processed -- connecting two points which are added to an ordered list, $O_1$. {\bf In panel 3}, more edges have been processed so that three ordered lists now exist. {\bf In panel 4}, ordered lists $O_1$ and $O_2$ have been merged to create $O_4$ -- the groups specified by $O_1$ and $O_2$ are also noted and stored within $G_\leq$ and $G_\geq$ respectively. {\bf In panel 5}, all edges have been processed and all points have been appended to ordered lists -- however multiple ordered lists still exist. {\bf In panel 6}, ordered lists $O_3$ and $O_4$ are merged to create $O_5$ -- as if there was an additional edge connecting the ordered lists (illustrated with a dashed blue line). Following this process, \astrolink\ has produced an ordered list, $O$ ($= O_5$ in this case), and a hierarchy of groups that embody the structure within the data set.}
    \label{fig:aggregation_visual}
\end{figure}

In \clustar\ this estimate of the density was used directly, however in \astrolink\ its logarithm is taken and re-scaled it between $0$ and $1$ such that
\begin{equation} \label{eq:scaledlogdensity}
    \log\hat{\rho}_i \coloneq \frac{\log(\rho_i/\rho_\mathrm{min})}{\log(\rho_\mathrm{max}/\rho_\mathrm{min})}.
\end{equation}
This transform renders all noisy fluctuations on the same scale regardless of their real density \citep[$\sigma_{\log\hat\rho}$ is approximately constant given fixed $k_\mathrm{den}$ and $d$;][]{Sharma2009} and it also allows us to see how large the affect of these fluctuations are compared to the range of densities within the data. For simplicity, we now refer to the set of the scaled (local) log-density values for all points as $\log\hat{\rho}$.

\subsection{Improving the Aggregation Process} \label{subsec:aggregation}

With a measure of local density computed, \astrolink\ now aggregates the data points together to form a hierarchy of groups. Conceptually (see Fig. \ref{fig:aggregation_visual} and the text below for specific algorithmic details), this process expands closed iso-density surfaces from high to low density. Whenever the iso-density value is lowered to that of a local density maxima, a new surface is born, and whenever a surface envelopes a new point, the point is appended to the end of an ordered list of the points located within that surface. We refer to the set of these ordered lists as $A$. As the iso-density value is lowered further, these surfaces (and their ordered lists) will merge together two at a time -- at which stage their start/end positions (as they appear within the newly-merged ordered list) are stored within $G_\leq$ and $G_\geq$ (for whichever list is smaller or larger respectively). Once this process has finished, all surfaces will have merged, one ordered list ($O$) will remain in $A$, and a hierarchy of all density-based groups can be retrieved from it by acquiring their start/end positions from the sets $G_\leq$ and $G_\geq$.

For \astrolink\ to be able to distinguish between \textit{clusters} and \textit{noise} in a way that is adaptable to varying levels of noise contamination, it can only identify groups as clusters \textit{after} it has found and compared each group of points in order to judge them as either more noise-like or more cluster-like (refer to Sec. \ref{subsec:identifyingclusters} for these details). This is unlike \clustar, which identifies clusters \textit{during} its aggregation process. Hence this change provides an opportunity to simplify and improve the effectiveness of the \astrolink\ aggregation process. 
%This process is shown in Fig. \ref{fig:aggregate}.

\subsubsection{Connectivity during the Aggregation Process} \label{subsubsec:connectivity}
Instead of actually constructing iso-density surfaces and checking which points become enveloped by them as the iso-density value is lowered, \astrolink\ relies upon the connections (referred to as edges from hereon) formed between nearest neighbours to decide which ordered list in $A$ a point should be appended to.

\citet{Oliver2022} found that by choosing $k_\mathrm{link} = \mathrm{max}\{\mathrm{ceil}(12.0d^{-2.2} - 23.0{k_\mathrm{den}}^{-0.6} + 10.0),\ 7\}$ as the number of nearest neighbours used for these edges, cluster resolution is maximised whilst algorithm run-time and the production of artificial disconnections during the aggregation process are minimised (refer to Sec. 6.1 of \citet{Oliver2022} for further details). Due to the similarities between \astrolink\ and \clustar, the first step of the aggregation process is therefore to reduce the nearest neighbour lists from $N_{k_\mathrm{den}}$ to $N_{k_\mathrm{link}}$. As such, the value of the $k_\mathrm{link}$ parameter will be automatically chosen as above ($k_\mathrm{link} = auto$, default) unless the user specifies otherwise.

\subsubsection{Aggregation}
In addition to using the edges between nearest neighbours to decide which ordered list in $A$ a point should be appended to, \astrolink\ also calculates a weight for each edge and uses these directly to decide the order in which each point is appended. The set of edges ($E$) is computed such that $E_{ij} = \mathrm{min}\{\log\hat{\rho}_i, \log\hat{\rho}_j\}$ where $p_i \in P$ and $p_j \in N_{k_\mathrm{link}, i}$, empty sets for $A$, $G_\leq$, $G_\geq$ are also initialised at this stage -- as shown in panel 1 of Fig. \ref{fig:aggregation_visual}. The edges are then processed in descending order such that for each edge, $E_{ij}$, an action is taken according to whether certain conditions are satisfied by the two points at the edge's vertices, $p_i$ and $p_j$, such that:
\begin{enumerate}[leftmargin = *, align = left]
    \item If neither point belongs to a list in $A$, create a new list with both points and add this to $A$. This is equivalent the birth of a new iso-density surface in the analogy above and is what has just occurred in panel 2 of Fig. \ref{fig:aggregation_visual}.
    %\item If both points belong to the same list in $A$, take no action.
    \item If exactly one point belongs to a list within $A$, add the other point to this list. This is equivalent to a new point becoming enveloped by an iso-density surface in the analogy above and is what has occurred for many points between panels 2 \& 3 as well as between panels 4 \& 5 of Fig. \ref{fig:aggregation_visual}.
    \item \label{action:iii} If both points belong to different ordered lists ($O_\geq$ and $O_\leq$, with sizes $n_\geq$ and  $n_\leq$) in $A$, then remove them from $A$, extend $O_\geq$ by $O_\leq$ to create $O_\mathrm{merged}$ and add it to $A$. Now add the start/end positions of $O_\geq$ and $O_\leq$ as they appear in $O_\mathrm{merged}$ to $G_\geq$ and $G_\leq$ respectively. This is equivalent to the merging of two iso-density surfaces in the analogy above and is what has just occurred in panel 4 of Fig. \ref{fig:aggregation_visual}.
\end{enumerate}
Notice that no action is taken if both points belong to the same list in $A$. In fact, the edges for which an action is taken (highlighted in each panel of Fig. \ref{fig:aggregation_visual}) are the same as those that belong to the minimum spanning \textit{forest} (not \textit{tree}, see below) that can be created using edge weights $E_{ij}' = f(E_{ij})$ -- where $f$ is some strictly decreasing function, e.g. $f(x) = 1/x$. In other words (and since the edges are considered sequentially in decreasing order), the aggregation process resembles Kruskal's minimum spanning tree (MST) \citep{Kruskal1956} algorithm interwoven with additional steps for tracking the hierarchy of groups that exist within the input data.

\begin{figure}
    \centering
    \includegraphics[trim={13.5mm 1mm 11mm 9.5mm}, clip, width=\columnwidth]{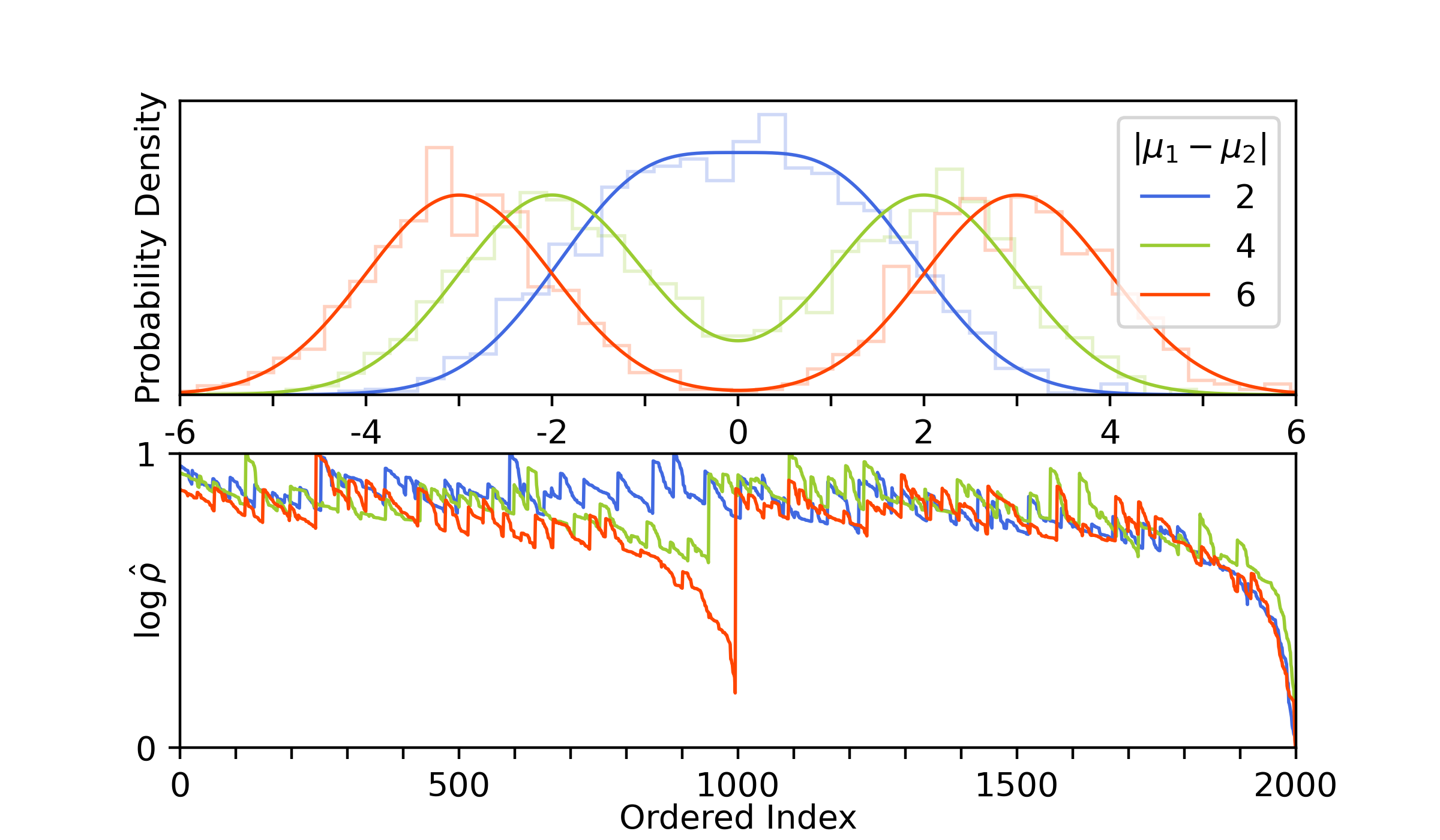}
    \vspace{-0.5cm}
    \caption{Examples of ordered-density plots (bottom) generated by \astrolink\ (using default settings) for three $1$-dimensional data sets that have been randomly sampled from two standard normal distributions with varying separations between them (top). We can see that as the separation grows, it becomes easier to distinguish the two distributions as individual clusters.}
    \label{fig:agg_example}
\end{figure}

However, it is not always possible to construct an MST from a set of edges unless they define a \textit{connected graph}. As \astrolink\ uses $n(k_\mathrm{link} - 1)$ edges (see Sec. \ref{subsubsec:connectivity}) to connect $n$ points, the edges will not always define a connected graph. The consequence of this is that after each of the edges in $E$ have been processed, multiple ordered lists may still remain in $A$. In this scenario, we imagine that there now exists some additional edges that connect pairs of points within each in the remaining ordered lists, so that the ordered lists are merged together in order of descending size -- this is illustrated in panel 6 of Fig. \ref{fig:aggregation_visual}. In effect, action \ref{action:iii} is repeated until a single ordered list exists in $A$ -- which from hereon we refer to as $O$.
%Notice that $A$ always contains groups that are still actively being appended to (i.e. active aggregations) and that for every saddle point in the density field the two active aggregations positioned \textit{uphill} from it are appended to $G$ (the smaller group, i.e. a subgroup) and $G_\mathrm{comp}$ (the larger group, i.e. the subgroup's complementary group). Separating the active aggregations into subgroups and their complementary groups helps to ensure an accurate description of the noisy density fluctuations -- as is detailed in Sec. \ref{subsec:identifyingclusters}. This process resembles Kruskal's minimum spanning tree (MST) \citep{Kruskal1956} interwoven with additional steps for tracking $P$'s subgroups and their hierarchy as it utilises connecting edges. However, instead of producing a minimal-weight tree it produces a maximal-weight forest since it uses density measures rather than distances to compute edge weights for $n(k_\mathrm{link} - 1)$ edges -- i.e. much less than $n(n - 1)$ edges. Following this, the largest of the remaining active aggregations is removed and used to begin the ordered list ($O$), then the others are processed in descending order of size so that for each $A_i \in A$; $A_i$ is appended to $G$, $O$ is appended to $G_\mathrm{comp}$, and then $A_i$ is appended to $O$. The aggregation process used by \astrolink\ is distinct from \clustar\ and helps it to increase the resolution of subgroups/clusters.

\subsubsection{The Ordered-Density Plot}
The ordered list, $O$, can be used to construct the \astrolink\ analogue of the \optics\ reachability plot -- the ordered-density plot. The ordered-density plot is a simple visualisation of the clustering structure as the complexity of the $n$-dimensional feature space has been reduced to the simplicity of a $2$-dimensional dendrogram. To construct the ordered-density plot, we need to compose the set of scaled log-densities with the ordered list which gives a function such that $f(i) = \log\hat{\rho}_i$, $\forall i \in O$. Fig. \ref{fig:agg_example} depicts the ordered-density plots for a series of input data sets consisting of two $1$-dimensional standard normal distributions at various separations.

The ordered-density plot reveals the input data points linked by decreasing density and based on their shared neighbourhood connectivity. Peaks in the plot represent overdensities as the points within them are both denser than their surrounds and ordered consecutively due to being locally connectable to each other. Fig. \ref{fig:agg_example} portrays two main peaks that become increasingly prominent as the separation between the distributions grows. Within these peaks are a series of smaller peaks that correspond to the stochastic clumping that arises whenever a data set contains spatial randomness -- as is the case here due to the data sets being created via random sampling. As the distribution separation increases, these smaller peaks become far less prominent than the two main peaks which indicates that a set of meaningful clusters is identifiable using the ordered-density plot.

\begin{figure}
    \centering
    \includegraphics[trim={13.5mm 1mm 11mm 9.5mm}, clip, width=\columnwidth]{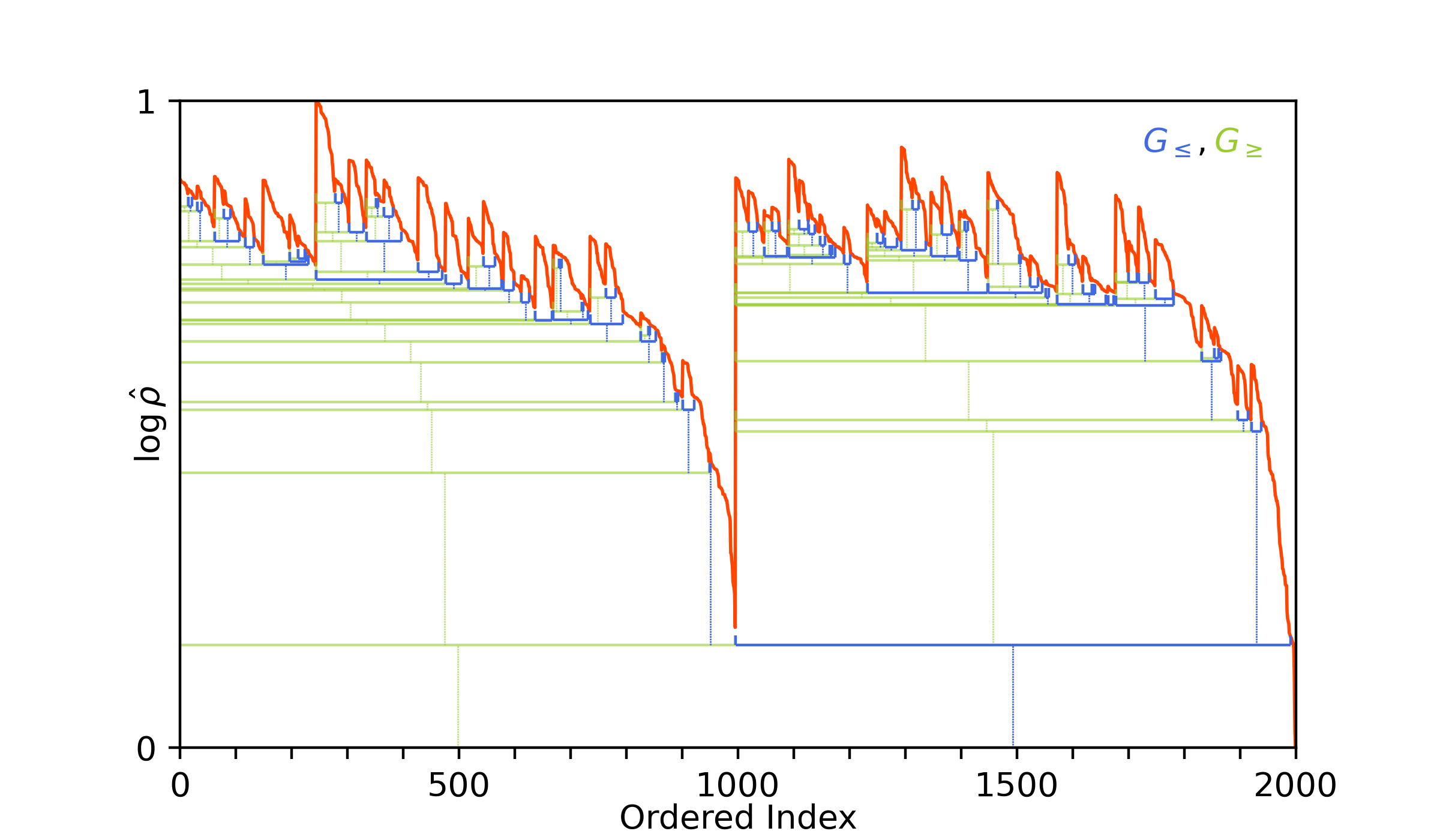}
    \vspace{-0.5cm}
    \caption{The \astrolink\ aggregation tree and its relation to the ordered-density plot for the $1$-dimensional data set with the largest separation from Fig. \ref{fig:agg_example}. The markings denote the groups, $G_\leq$ (blue) and $G_\geq$ (semi-transparent yellow-green), found during the aggregation process.}
    \label{fig:merg_example}
\end{figure}

\subsubsection{The Aggregation Tree} \label{subsubsec:aggregationtree}
A hierarchy of groups, defined by the sets $G_\leq$ and $G_\geq$, is also generated during the aggregation process -- an example of which is shown in Fig. \ref{fig:merg_example}. For every group denoted in $G_\leq$ there is exactly one (larger) group denoted in $G_\geq$, as these two groups are merged whenever a saddle point in the density field is found during the aggregation process -- i.e. whenever action \ref{action:iii} is performed. As such, the groups in $G_\leq$ are smaller and are generally less prominently peaked within the ordered-density plot than their counterparts from $G_\geq$. Hence, the groups in $G_\leq$ tend to be more noise-like than those in $G_\geq$. This makes any true clusters within $G_\leq$ appear more distinct from the other groups in $G_\leq$ than any true clusters within $G_\geq$ do when compared to the other groups in $G_\geq$.

A hierarchy of clusters determined from $G_\leq$ alone is structurally akin to that produced by the \code{SUBFIND} \citep{Springel2001} and \code{EnLink} \citep{Sharma2009} algorithms. While this is suitable for some clustering scenarios, as evident in Fig. \ref{fig:merg_example}, finding clusters from $G_\leq$ alone can only capture one of the two standard normal distributions. Thus, in many scenarios, the hierarchy from $G_\geq$ is also needed in order to identify all relevant clusters.

\subsection{Identifying Clusters} \label{subsec:identifyingclusters}
\astrolink\ now identifies clusters by assessing the statistical distinctiveness of groups from the aggregation tree within the ordered-density plot. By measuring the \textit{clusteredness} of each group and fitting a model to the distribution of this measure for all groups in $G_\leq$, groups can then be classified as inliers (noise) or outliers (clusters) from this distribution. An optional hierarchy correction can then also be performed by incorporating the groups in $G_\geq$.

\subsubsection{The Clusteredness of a Group}
\begin{figure}
    \centering
    \includegraphics[trim={13.5mm 0mm 11mm 9.5mm}, clip, width=\columnwidth]{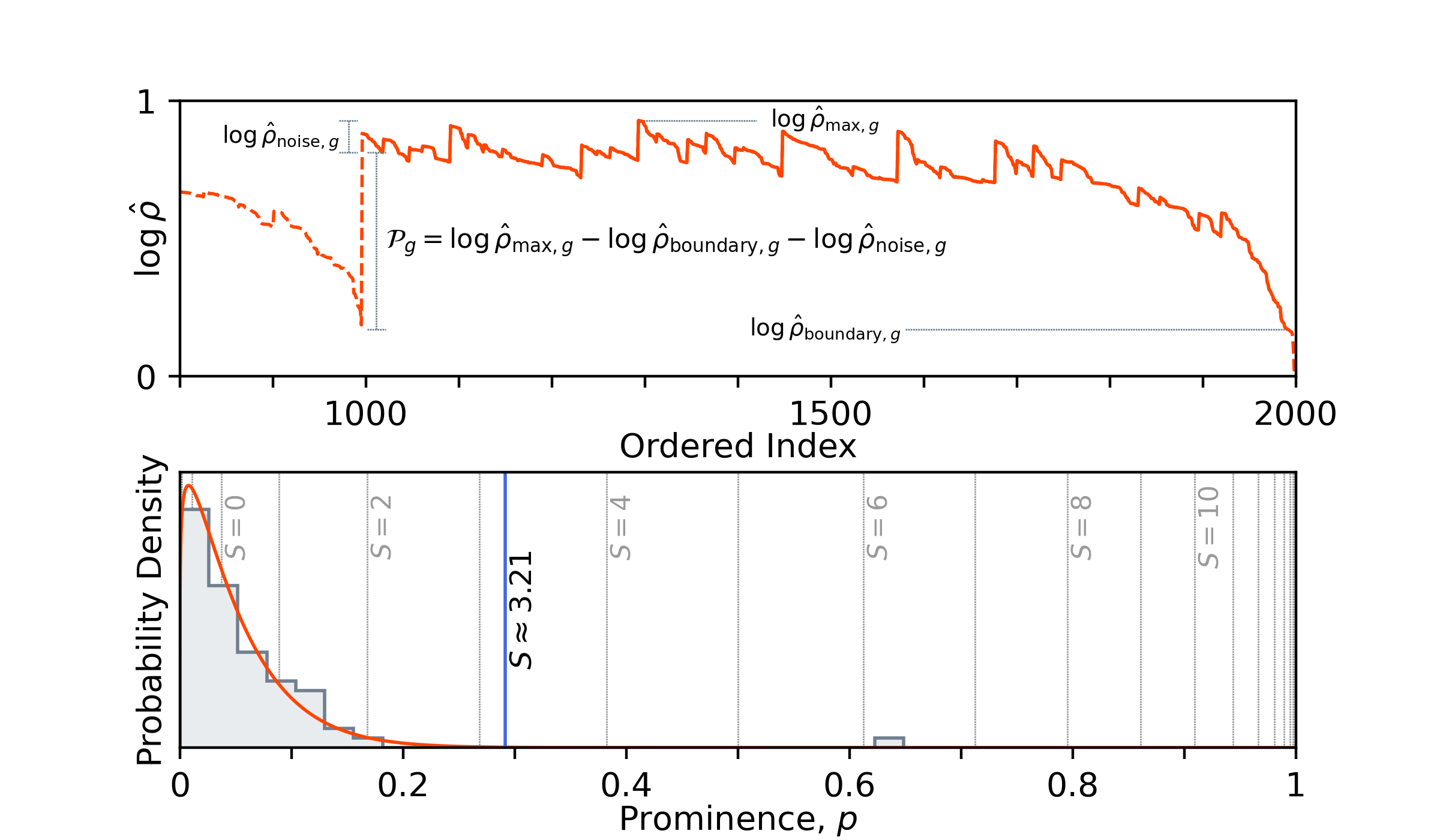}
    \vspace{-0.5cm}
    \caption{An illustration of the prominence calculation (top) for the largest group in $G_\leq$ and the resulting prominence distribution (bottom) for all such groups (groups are shown in Fig. \ref{fig:merg_example}). The largest group is objectively more clustered, and hence is a clear outlier ($S > 6$), from the others. The solid red line indicates the model for group prominences (detailed in Sec. \ref{subsubsec:prominencemodel}) and the solid blue line, positioned at $p = c$, is labelled with the corresponding data-driven estimate of $S$ (detailed in Sec. \ref{subsubsec:significance}).}
    \label{fig:prom_dist_example}
\end{figure}

Intuitively, the measure of a group's \textit{clusteredness} should consider its overdensity and also account for the magnitude of intra-group noise. Due to the logarithmic scaling of densities from Eq. \ref{eq:scaledlogdensity}, vertical differences on the ordered-density plot represent the logarithm of a ratio of densities. As such, we can calculate the logarithm of a group's overdensity factor by finding the difference between some characteristic $\log\hat\rho$ value of that group and the $\log\hat\rho$ value of its surrounds. The $\log\hat\rho$ value \textit{at its surrounds} is taken as the $\log\hat\rho$ saddle-point that was used to merge the group into the aggregation tree -- which is the maximum $\log\hat\rho$ that can be found at the group's boundary. The group's characteristic $\log\hat\rho$ value is taken to be maximum $\log\hat\rho$ found within the group minus a measure of the intra-group noise. As such, we construct a statistical measure of clusteredness, which we call the \textit{prominence}, such that for any group, $g$;
\begin{equation} \label{eq:prominence}
    \mathcal{P}_g = \log\hat{\rho}_{\mathrm{max}, g}% \notag \\
    - \log\hat{\rho}_{\mathrm{boundary}, g}% \notag \\
    - \log\hat{\rho}_\mathrm{noise, g}.% \notag \\
%    =&\ \mathrm{max}\{\log\hat{\rho}_i \mid \forall p_i \in g\} \notag \\
%    -&\ \mathrm{max}\{\log\hat{\rho}_j \mid p_j \in N_{k_\mathrm{link}, i} \cap g,\ \forall p_i \in g\} \notag \\
%    -&\ \sqrt{\frac{1}{N_\mathrm{child}}\sum\left(\log\hat{\rho}_{\mathrm{max}, \mathrm{child}} - \log\hat{\rho}_{\mathrm{boundary}, \mathrm{child}}\right)^2}.
\end{equation}
Here $\log\hat{\rho}_{\mathrm{max}, g}$ is the maximum value of $\log\hat{\rho}$ within $g$, $\log\hat{\rho}_{\mathrm{boundary}, g}$ is the value of $\log\hat{\rho}$ at the saddle point that connects that $g$ to another, and $\log\hat{\rho}_\mathrm{noise, g}$ is a term that accounts for intra-group noise. The noise-correction term is calculated as the root-mean-square of the $\log\hat{\rho}_{\mathrm{max}, \mathrm{child}} - \log\hat{\rho}_{\mathrm{boundary}, \mathrm{child}}$ values for each of the direct children (excluding children of children) groups of $g$. As such, this measure may be viewed as the logarithm of a signal to noise ratio. The top panel of Fig. \ref{fig:prom_dist_example} shows a visualisation of the calculation of $\mathcal{P}_g$ for the largest group from $G_\leq$ as shown in Fig. \ref{fig:merg_example}.

\subsubsection{A Model for Group Prominences} \label{subsubsec:prominencemodel}
\astrolink\ now fits a descriptive model, $M$, on a case-by-case basis to the prominences of the groups from $G_\leq$\footnote{Including the prominences of the groups $G_\geq$ only serves to add weight to the tail of the prominence distribution -- making it more difficult to identify outliers (clusters).} by minimising the corresponding negative log-likelihood. For density estimates of Poisson noise computed with a fixed number of nearest neighbours, the probability distribution of $\rho$ should be approximately Gaussian \citep{Sharma2009}. As such, the prominence (which is effectively an absolute difference of these estimates) should belong to a half-normal distribution. With real data however, some critical assumptions are violated; i.e. generally the nearest neighbours are not drawn from homogeneous Poisson noise; the density estimation and aggregation processes impose small-scale smoothing effects on the density field that suppress the production of groups with fewer points than $N_{k_\mathrm{den}}$ and $N_{k_\mathrm{link}}$ respectively; etc. As such, we construct $M$ to be a combination of two probability distributions -- one for noise and one for clusters -- such that $M(p) = H(c - p)M_\mathrm{noise}(p) + H(p - c)M_\mathrm{clusters}(p)$ where $p$ is a random variable representing prominence, $H(x)$ is the Heaviside step function, and $c \in [0, 1]$ is a parameter of the model.

For $M_\mathrm{noise}$, we have used the second-order correction estimate of the Akaike information criterion \citep[AIC \& AICc;][]{Akaike1974, Hurvich1989} and the Bayesian information criterion \citep[BIC;][]{Schwarz1978} to assess the suitability of the Log-Normal, Inverse Gaussian, Gamma, Beta, Beta-Prime, Generalised Gamma, and Generalised Beta-Prime distributions when \astrolink\ is applied to a range of $n$-point $d$-dimensional uniform distributions on the unit hypercube (for various combinations of $n$ and $d$). With both criteria, we found that using the Beta distribution for $M_\mathrm{noise}$ is best (with parameters $a, b \geq 1$ and typically $b \gg a$, i.e. uni-modal with a positively-skewed tail). We also found that using a Uniform distribution for $M_\mathrm{clusters}$ provides enough shape flexibility so that $M_\mathrm{noise}$ remains largely unaffected by the presence of outliers. This can be seen in the bottom panel of Fig. \ref{fig:prom_dist_example} as the presence of a $S > 6$ outlier does not detract from the fitting-quality of $M_\mathrm{noise}$. Using the model parameters ($a, b, c$), $M_\mathrm{noise}(p)$ and $M_\mathrm{clusters}(p)$ are also weighted so that $M(p)$ is normalised and continuous on the interval $p \in [0, 1]$.

\begin{figure}
    \centering
    \includegraphics[trim={13.5mm 1mm 11mm 9.5mm}, clip, width=\columnwidth]{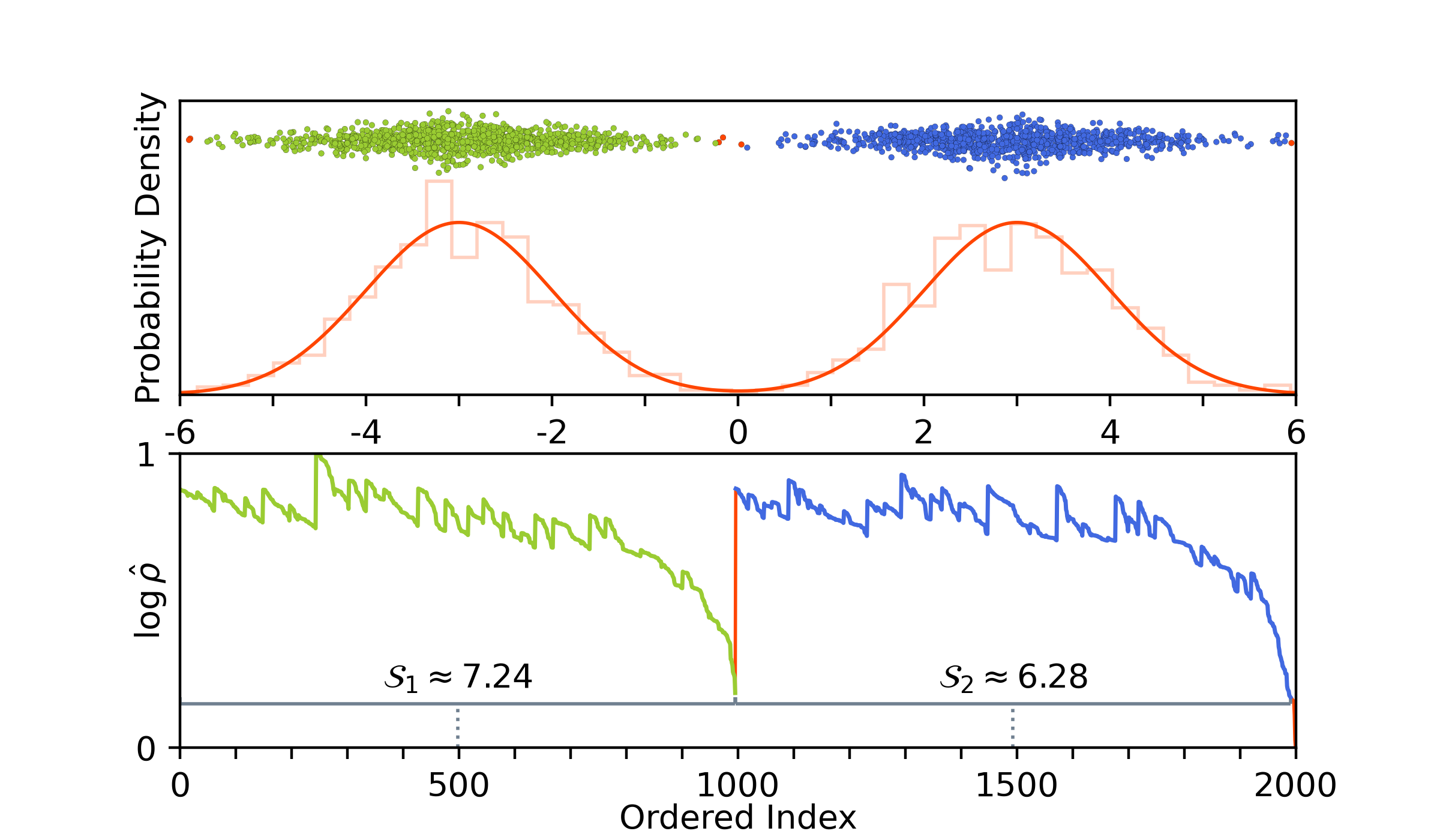}
    \vspace{-0.5cm}
    \caption{Clusters identified from the toy example of Fig. \ref{fig:agg_example} using \astrolink\ with default settings. Shown here is the $1$-dimensional probability distribution alongside the input data that was sampled from it (top) and the ordered-density plot, the final hierarchy tree, and the clusters' significances (bottom). This data is coloured according to the cluster that it has been allocated to by \astrolink. \astrolink\ has differentiated the two Gaussian distributions from the noisy density fluctuations within them with a statistical significance of $7.24\sigma_\mathrm{noise}$ and $6.28\sigma_\mathrm{noise}$ respectively.}
    \label{fig:final_clusters}
\end{figure}

\subsubsection{The Statistical Significance of Clusters} \label{subsubsec:significance}
To label groups in $G_\leq$ as clusters, \astrolink\ now simply identifies all such groups that have a statistical significance ($\mathcal{S}$) that is greater than $S\sigma_\mathrm{noise}$. To be clear, any group ($g \in G_\leq$) will now be labelled as a cluster if its prominence ($\mathcal{P}_g$) satisfies
\begin{equation} \label{eq:significance}
    \mathcal{S}_g \coloneq {F_{\mathcal{N}(0, 1)}}^{-1} \left[F_{\beta(a, b)}(\mathcal{P}_g) \right] \geq S,
\end{equation}
where $F_{\mathcal{N}(0, 1)}$ and $F_{\beta(a, b)}$ are the standard normal and Beta cumulative distribution functions respectively. The parameters $a$ and $b$ are derived by fitting the model outlined in Sec. \ref{subsubsec:prominencemodel} to the distribution of group prominences. With this, the group prominences are transformed so that the group significances follow a standard normal with a long positive tail of outliers containing the significances of increasingly more clustered groups. The $S$ parameter is therefore a measure of how clustered -- relative to the noise present within the data -- the resultant clusters are. The value of $S$ may be chosen by the user but the default value is $S = auto$, in which case \astrolink\ calculates it according to the model parameter, $c$, i.e. $S = {F_{\mathcal{N}(0, 1)}}^{-1} \left[F_{\beta(a, b)}(c) \right]$. Since $c$ marks the prominence value for which the groups transition from noise to clusters, it can therefore be used to automatically estimate an appropriate value for $S$.

\subsubsection{Correcting the Hierarchy} \label{subsubsec:correction}
To ensure all relevant clusters are included in the final hierarchy, \astrolink\ can optionally label the groups from $G_\geq$ as clusters as well as those from $G_\leq$. If {\tt h\_style} $ = 1$ (default), then for each cluster already found, it's corresponding sibling group in $G_\geq$ is considered if it also satisfies Eq. \ref{eq:significance}. Then for each of these cluster-like sibling groups in $G_\geq$, \astrolink\ labels it a cluster if it is the smallest such group in $G_\geq$ that shares its starting position within the ordered list, $O$. The latter condition simplifies and reduces the depth of the resultant hierarchy by eliminating clusters-within-clusters that only differ by a small number of points. With these additional clusters the resultant hierarchy is similarly styled to \clustar, however now each cluster is a more meaningful and statistically interpretable group. Fig. \ref{fig:final_clusters} now depicts the final hierarchy of clusters extracted from the same toy data set shown in Figs. \ref{fig:agg_example} -- \ref{fig:prom_dist_example}. This step can be ignored by setting {\tt h\_style} $= 0$.

\section{Performance Analysis} \label{sec:performance}
We now assess the performance of \astrolink\ by applying it to a series of synthetic galaxies each with a set of ground-truth labels. We analyse the time and space complexity of \astrolink\ as well as it's ability to produce a meaningful set of astrophysical clusters in comparison to its algorithmic predecessors -- \clustar\ and \hoptics. Lastly, we show some of these outputs visually.

\subsection{Synthetic Data}
To assess \astrolink's performance we use \code{Galaxia} \citep{Sharma2011} to produce random samples of the $11$ $\Lambda$CDM stellar haloes from \citet{Bullock2005} and the complementary $6$ artificial stellar haloes from \citet{Johnston2008}. Each of the original $\Lambda$CDM stellar haloes are simulated using a hybrid semi-analytical and hydrodynamic $N$-body approach. Satellites are first modelled as $N$-body dark matter systems within a parent galaxy whose disk, bulge, and halo are defined by time dependent semi-analytic functions. The simulation uses a $\Lambda$CDM cosmology with parameters $\Omega_m = 0.3$, $\Omega_\Lambda = 0.7$, $\Omega_bh^2 = 0.024$, $h = 0.7$, and $\sigma_8 = 0.9$. Semi-analytical models then assign stellar populations to each satellite while a chemical enrichment model \citep{Robertson2005, Font2006} calculates age-appropriate metallicities for the star particles. This process yields satellites whose structural properties are in agreement with the Local Group's dwarf galaxies.

With this regime each satellite has three main model parameters; the time since accretion ($t_{\mathrm{acc}}$), the luminosity ($L_{\mathrm{sat}}$), and the orbital circularity ($\epsilon = J/J_{\mathrm{circ}}$). The distribution of these parameters specify the accretion history of a halo and as such a further $6$ artificial haloes were created in \citet{Johnston2008} for the purpose of studying the effects of different accretion histories on the properties of haloes. These $6$ artificial haloes are characterised by accretion histories that are predominantly radial ($\epsilon < 0.2$); circular ($\epsilon > 0.7$); old ($t_{\mathrm{acc}} > 11$ Gyr); young ($t_{\mathrm{acc}} < 8$ Gyr); high luminosity ($L_{\mathrm{sat}} > 10^7 L_{\mathrm{\odot}}$); and low luminosity ($L_{\mathrm{sat}} < 10^7 L_{\mathrm{\odot}}$).

For these galaxies, \code{Galaxia} provides spatial (${\bf x} = (x, y, z)$), kinematic (${\bf v} = (v_x, v_y, v_z)$), and chemical (${\bf m} = ($[Fe/H]$, $[$\alpha$/Fe]$)$) quantities for each particle, along with ground-truth labels indicating the particle's satellite of origin and also whether each satellite is self-bound. More details on these simulations can be found in Sec. 3.4 of \citet{Sharma2011} and references therein. With this information, we apply \astrolink\ to the various galaxies and assess its performance.

\subsection{Time Complexity}
\begin{figure}
    \centering
    \includegraphics[trim={5mm 5mm 11mm 15mm}, clip, width=\columnwidth]{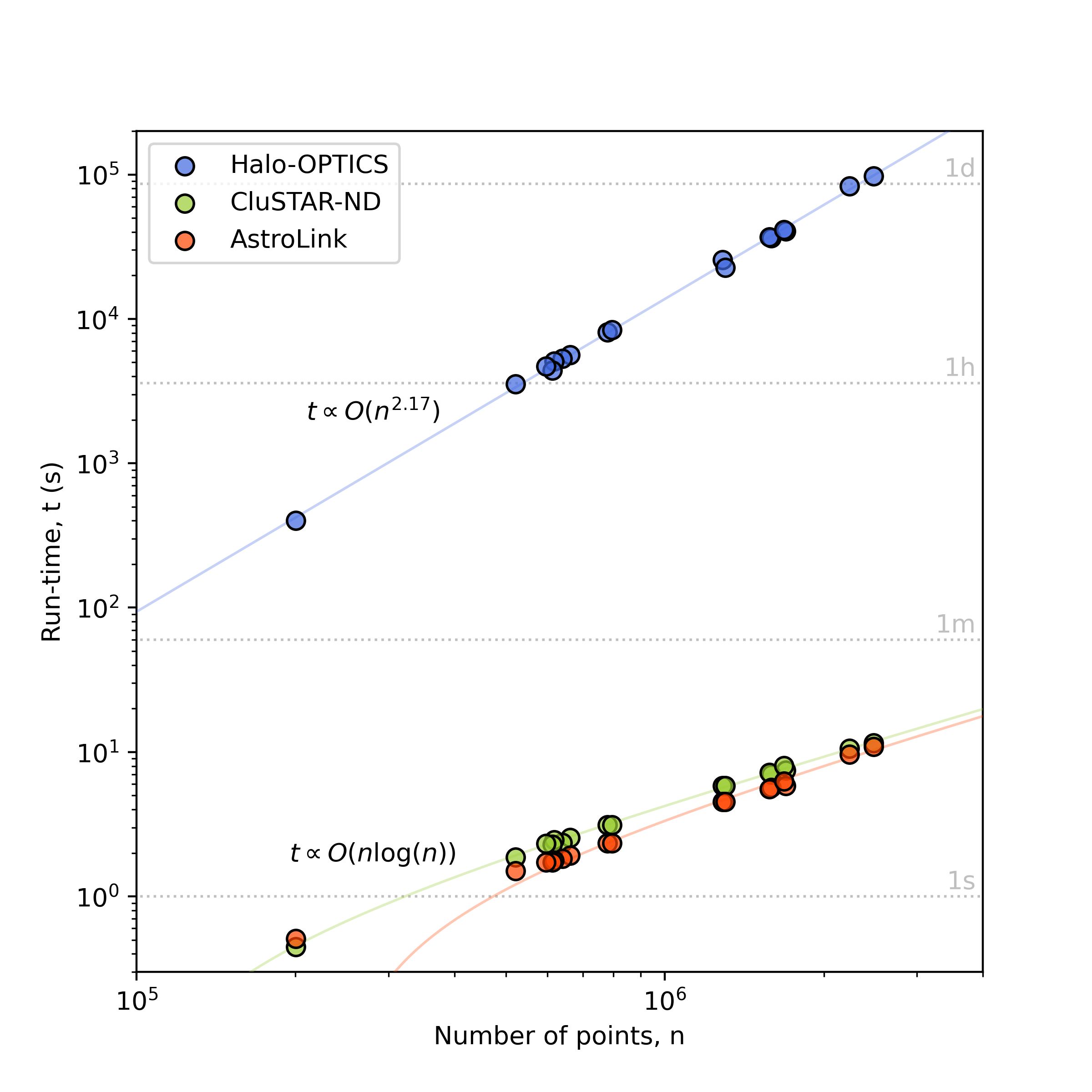}
    \vspace{-0.5cm}
    \caption{The run-times and time-complexities of \astrolink, \clustar, and \hoptics\ when applied to the spatial positions ($\bf{x}$) of particles within the synthetic galaxies from \code{Galaxia}. The curves of best fit correspond to the time complexities for the algorithms. \astrolink\ and \clustar\ both exhibit a run-time improvement of many orders of magnitude over \hoptics, and even though their time complexities are the same, \astrolink's smaller constant factor means that (asymptotically) it is an additional $\sim3$ times faster than \clustar. Furthermore, since \astrolink\ may be run in parallel for a larger portion of its computation, it is the most-suitable of the three algorithms for applications to large data sets.}
    \label{fig:runtimescomparison}
\end{figure}

We now compare the run-time of \astrolink\ to that of \clustar\ and \hoptics. Fig. \ref{fig:runtimescomparison} shows the run-times of the three algorithms when applied to the spatial positions of the synthetic galaxies from \code{Galaxia}. All runs were performed using a single core of a $13^\mathrm{th}$ Gen Intel i7-13700HX processor with a clock-speed of $5.00$ GHz and all codes are written using \code{Python3} -- each making use of optimised numerical packages such as \code{NumPy} \citep{Numpy}, \code{SciPy} \citep{SciPy}, \code{Scikit-learn} \citep{scikit-learn}, and \code{Numba} \citep[\astrolink\ \& \clustar\ only;][]{Numba} to differing degrees.

It is clear from Fig. \ref{fig:runtimescomparison} that \astrolink\ outperforms its predecessors in terms of running time. The time complexities of \astrolink\ and \clustar\ are $\mathcal{O}(n\mathrm{log}(n))$ while \hoptics\ exhibits super-quadratic time complexity at $\mathcal{O}(n^{2.17})$. For \hoptics, this results from a combination of it performing a radial search about every data point and that the cuspyness of the haloes increase with their size -- this is discussed further in \citet{Oliver2022}. This does not impact \astrolink\ or \clustar, as they only perform a $k_\mathrm{den}$ nearest neighbour search for each data point. While this $k_\mathrm{den}$ nearest neighbour search still contributes the largest portion of their total run-time. Comparatively, \astrolink\ is approximately $1.27$ times faster than \clustar\ and between $2400$--$9000$ times faster than \hoptics\ over these data sets.

These run-times can be further improved by using multiple cores in parallel. For \astrolink, all steps are parallelised over shared memory -- except the part of the aggregation process resembling Kruskal's MST. However, the majority of the computation in this step can also be parallelised \citep{Bader2006} by combining Prim's MST algorithm \citep{Jarnik1930, Prim1957, Dijkstra2022} and Boruvska's MST algorithm \citep{Boruuvka1926,Choquet1938, Sollin1965} -- an implementation we leave as future work.

\subsection{Space Complexity}
In terms of memory footprint, each algorithm has $O(n)$ space complexity, although with differing constant factors. \clustar\ has the largest of these factors since it stores lists of indices for each cluster, while \astrolink\ and \hoptics\ compute an ordered list and then only store the start and end positions of each cluster within this list. Conversely, \hoptics\ has the smallest constant factor, as it only stores indices of the radial nearest neighbours for one point at a time -- whereas \astrolink\ and \clustar\ keep copies of the $k_\mathrm{den}$ nearest neighbours of every point. Consequently, \astrolink\ boasts the median constant factor and thus it presents a trade-off between time and space complexity while improving interpretability and reliability.

\begin{figure}
    \centering
    \includegraphics[trim={11mm 54mm 14mm 36mm}, clip, width=\columnwidth]{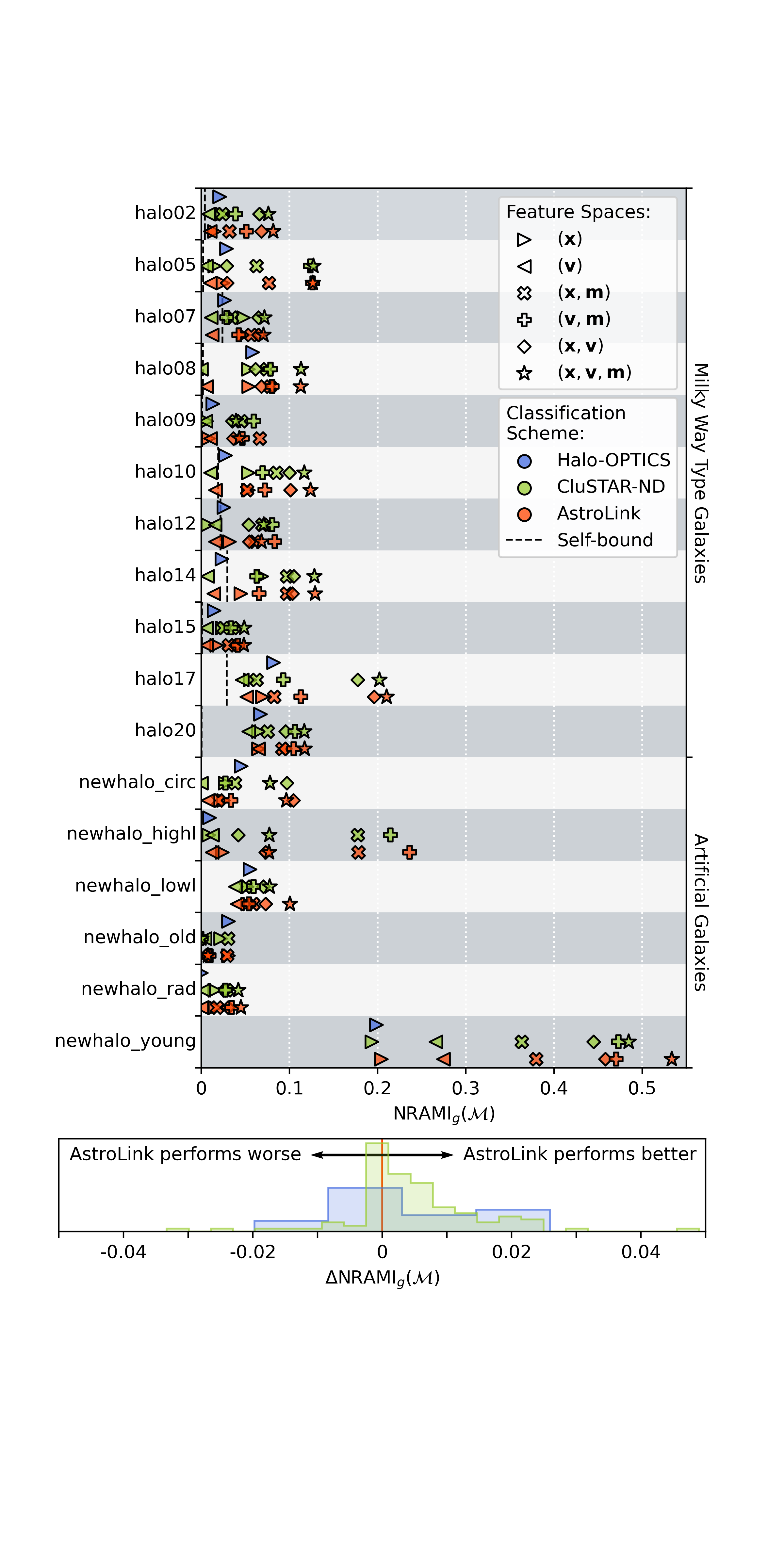}
    \vspace{-0.5cm}
    \caption{The NRAMI between for the \astrolink, \clustar, and \hoptics\ algorithms when applied to each synthetic galaxy and feature space combination (top) and the differences in the NRAMI between the \astrolink\ algorithm and its predecessors for each of these clustering scenarios (bottom). The NRAMI of a hypothetical set of clusterings that exactly classify the self-bound satellites within each MW type galaxy are also shown in the top panel as a reference. \astrolink\ outperforms \clustar\ in $\sim75\%$ of comparisons and \hoptics\ in $\sim50\%$ of comparisons.}
    \label{fig:finalmutualinfo}
\end{figure}

\subsection{Clustering Power}
We now assess the capability of \astrolink\ to retrieve a set of astrophysical clusters in comparison to \clustar\ and \hoptics. We first define an information-based statistic that represents the portion of meaningful classification that has been made by the algorithms, we then run each algorithm over the synthetic galaxies, before finally comparing their outputs.

\subsubsection{Measuring Clustering Power}
We use the normalised random-adjusted mutual-information presented (NRAMI) in Sec. 5.1.1 of \citet{Oliver2022} to assess the goodness-of-fit between a clustering ($C_g$) produced by \astrolink/\clustar/\hoptics\ and the ground-truth labels ($T_g$) provided by \code{Galaxia} for a given galaxy $g$. To prepare the hierarchy of clusters from these algorithms for comparison to the flat ground-truth labels, we first assign a unique label to every point of the input data that corresponds to the smallest cluster that it is predicted to be a part of -- we refer to this flattened set of labels as $C_g^*$. The knowledge gained about a galaxy's satellites from a clustering algorithm can then be quantified by the mutual information \citep{Shannon1948} between $C_g^*$ and $T_g$, such that
\begin{align} \label{eq:mutualinformation}
    &I(C_g^*; T_g) = H(T_g) - H(C_g^*|T_g),\ \mathrm{where} \notag\\
    &H(T_g) \equiv -\sum_{t \in T_g} P(t)\ \mathrm{log}(P(t)),\ \mathrm{and} \notag\\
    &H(C_g^*|T_g) \equiv -\sum_{c \in C_g^*,\ t \in T_g} P(c, t)\ \mathrm{log}\bigg(\frac{P(c, t)}{P(t)}\bigg).
\end{align}
Here $H(T_g)$ is the entropy of $T_g$ and $H(C_g^*|T_g)$ is the conditional entropy of $C_g^*$ given $T_g$. The portion of \textit{relevant} information learnt can then be calculated by normalising this to ensure that the expected knowledge gained about $T_g$ from a random re-clustering of $C_g^*$ (i.e. a set of randomly assigned labels with equal sizes as those in $C_g^*$) is represented by a value of $0$, and that similarly, the knowledge gained from a perfect clustering ($C_g^* = T_g$) is represented by a value of $1$. Hence, if $\mathcal{M}$ represents some feature space combination upon which the predicted clusters are dependent, the measure we use to assess an algorithm's clustering power is defined as
\begin{equation} \label{eq:objectivefunction}
    \mathrm{NRAMI}_g(\mathcal{M}) = \frac{I(C_g^*(\mathcal{M}); T_g) - E[I(R_g^*(\mathcal{M}); T_g)]}{H(T_g) - E[I(R_g^*(\mathcal{M}); T_g)]},
\end{equation}
where $E[I(R_g^*(\mathcal{M}); T_g)]$ is the expected mutual information between a random re-clustering, $R_g^*(\mathcal{M})$, and $T_g$. Refer to Sec. 5.1.1 of \citet{Oliver2022} for more details on this measure.

\subsubsection{Comparison}
We now compare each algorithm's clustering power by applying them to each synthetic galaxy over various combinations of the available feature space. We select $\mathcal{M}$ from the combinations of the spatial (${\bf x}$), kinematic (${\bf v}$), and chemical (${\bf m}$) feature spaces, i.e. $({\bf x})$, $({\bf v})$, $({\bf x}, {\bf m})$, $({\bf v}, {\bf m})$, $({\bf x}, {\bf v})$, and $({\bf x}, {\bf v}, {\bf m})$. We apply \clustar\ and \hoptics\ using their default settings \citep[refer to][]{Oliver2022}, but since their cluster extraction parameters have been optimised for these galaxies we alter the significance threshold parameter, $S$, of \astrolink\ in this comparison.

For values of $S$ in the range of $\sim3$ to $\sim6$, the NRAMI between the \astrolink\ output and the ground-truth hovers around values that are consistent with the \clustar\ output. By default ($S = auto$), the data-driven value of $S$ falls within this range for all clustering scenarios -- and so naively one can expect to achieve a roughly equal clustering power when using \astrolink's default settings. However since the output of \astrolink\ is more easily interpretable and the significance parameter is intuitive to adjust, we set the $S$ parameter on a case-by-case basis in order to produce the best possible value of the $\mathrm{NRAMI}_g(\mathcal{M})$. This allows us to see what \astrolink\ is realistically capable of achieving in a practical use-case/data-exploration scenario.

Fig. \ref{fig:finalmutualinfo} depicts the clustering performance of each algorithm when applied to each galaxy and feature space combination as well as the difference in clustering power for each algorithm. Here \astrolink\ is shown to generally outperform the other algorithms, and even when it does not, the difference is most-often negligible. It is also important to note that \astrolink\ is effectively being penalised by the NRAMI measure -- since it tends to produce a shallower hierarchy than the other two codes. With fewer levels to the hierarchy there are fewer groupings at high densities which improves the interpretability of the \astrolink\ output but in this case imposes a penalty to \astrolink\ in this comparison. Within this figure, it can also be noted that if the feature space of the input data has a larger number of more informative dimensions then the clustering power will generally be improved. Additionally, the type of clusters within the galaxy has a large effect on the clustering power -- young or high luminosity satellites are well-matched whereas those that are old or that are orbiting on predominantly radial trajectories are more difficult to recover.

\subsection{Visualising Clustering Structure}
\begin{figure}
    \centering
    \includegraphics[trim={12mm 23mm 9mm 36mm}, clip, width=\columnwidth]{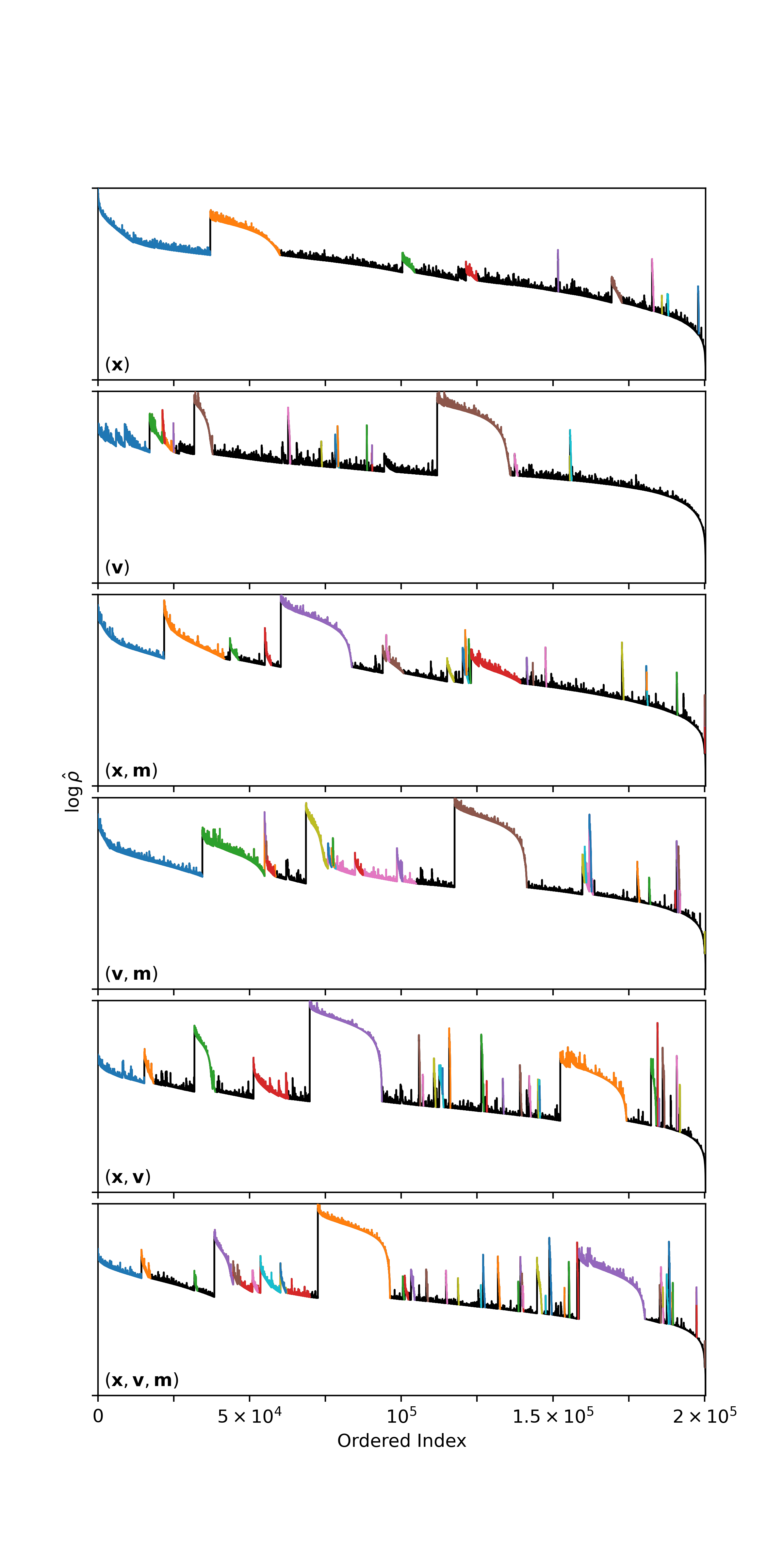}
    \vspace{-0.5cm}
    \caption{Ordered density plots produced by \astrolink\ when applied to the artificial galaxy, newhalo\_young, over various feature space combinations. Clusters are represented with a colour cycle of nine colours (also depicted in Fig. \ref{fig:newhalo_young_3d_visuals}) and have been identified with \astrolink's default settings (unlike in Fig. \ref{fig:finalmutualinfo}, $S$ is now estimated from the data)}.
    \label{fig:newhalo_young_ordered_density}
\end{figure}

We now also visually present the \astrolink\ output in order to further demonstrate its performance and usability. The ordered-density plot produced by \astrolink\ holds the information attainable from the estimated density field\footnote{The information attainable from the density field can be enhanced by first computing a locally adaptive metric before the density estimation step and/or by applying some post-process algorithm that meaningfully attributes additional points to the clusters predicted by \astrolink.}, so in Fig. \ref{fig:newhalo_young_ordered_density} we illustrate how this information varies for the predominantly young artificial galaxy from \citet{Johnston2008} when using different feature spaces to find the clustering of the data. Here it can be seen that the complexity of the clustering structure increases with the number of informative dimensions and hence visualising the ordered-density plot provides valuable insight to the user. Furthermore, the ordered-density plot can be used as a guide to the user when fine-tuning the significance threshold ($S$) -- as unclassified structure can be classified by lowering $S$ and, \textit{vice versa}, classified noise can be declassified by raising $S$.

\begin{figure}
\vspace{-0.14cm}
\begin{center}
\begin{tikzpicture}
  \node (img1) {\includegraphics[height=4.7cm]{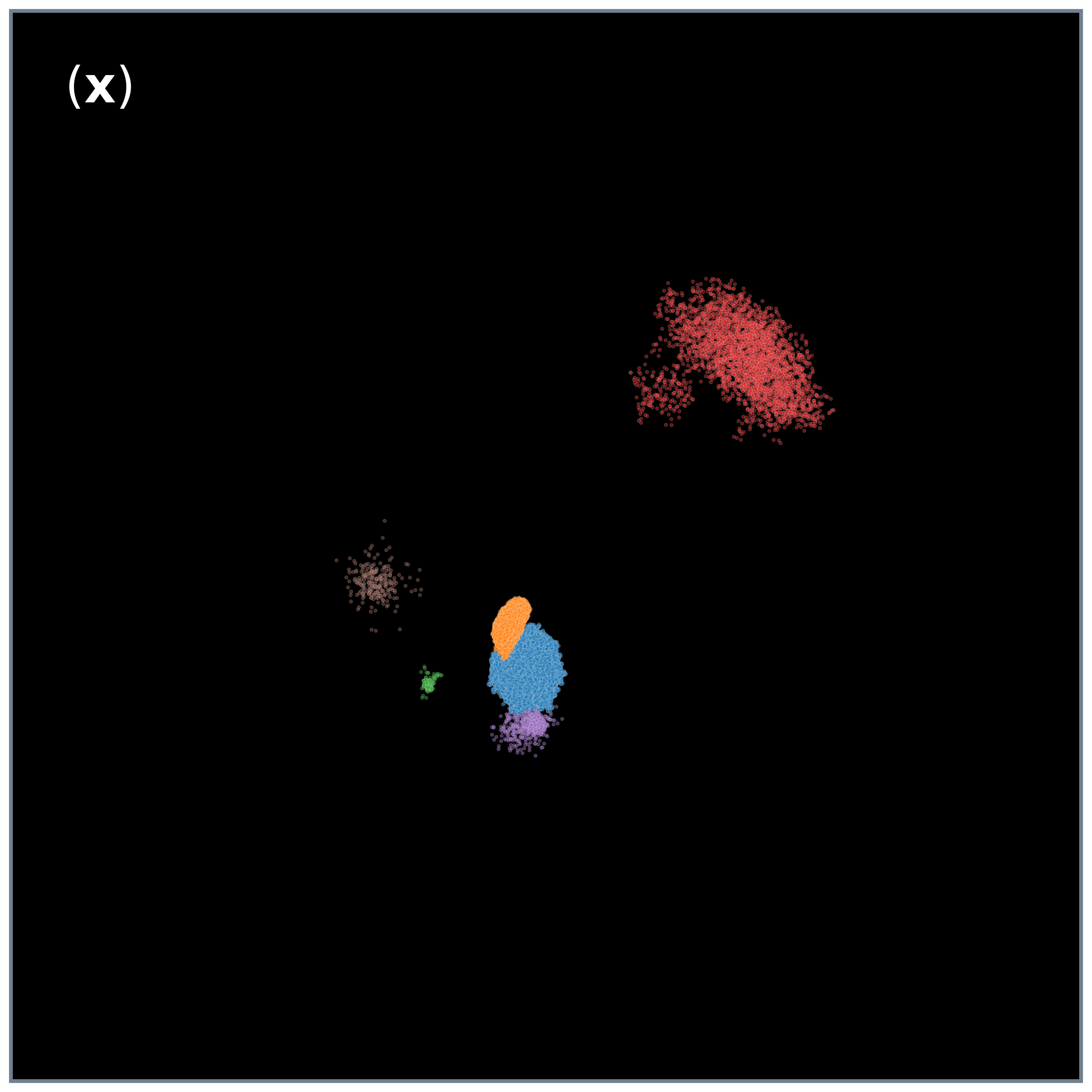}};
  %\pause
  \node (img2) at (img1.south east) [xshift = 1.25cm, yshift = 1.3cm] {\includegraphics[height=4.7cm]{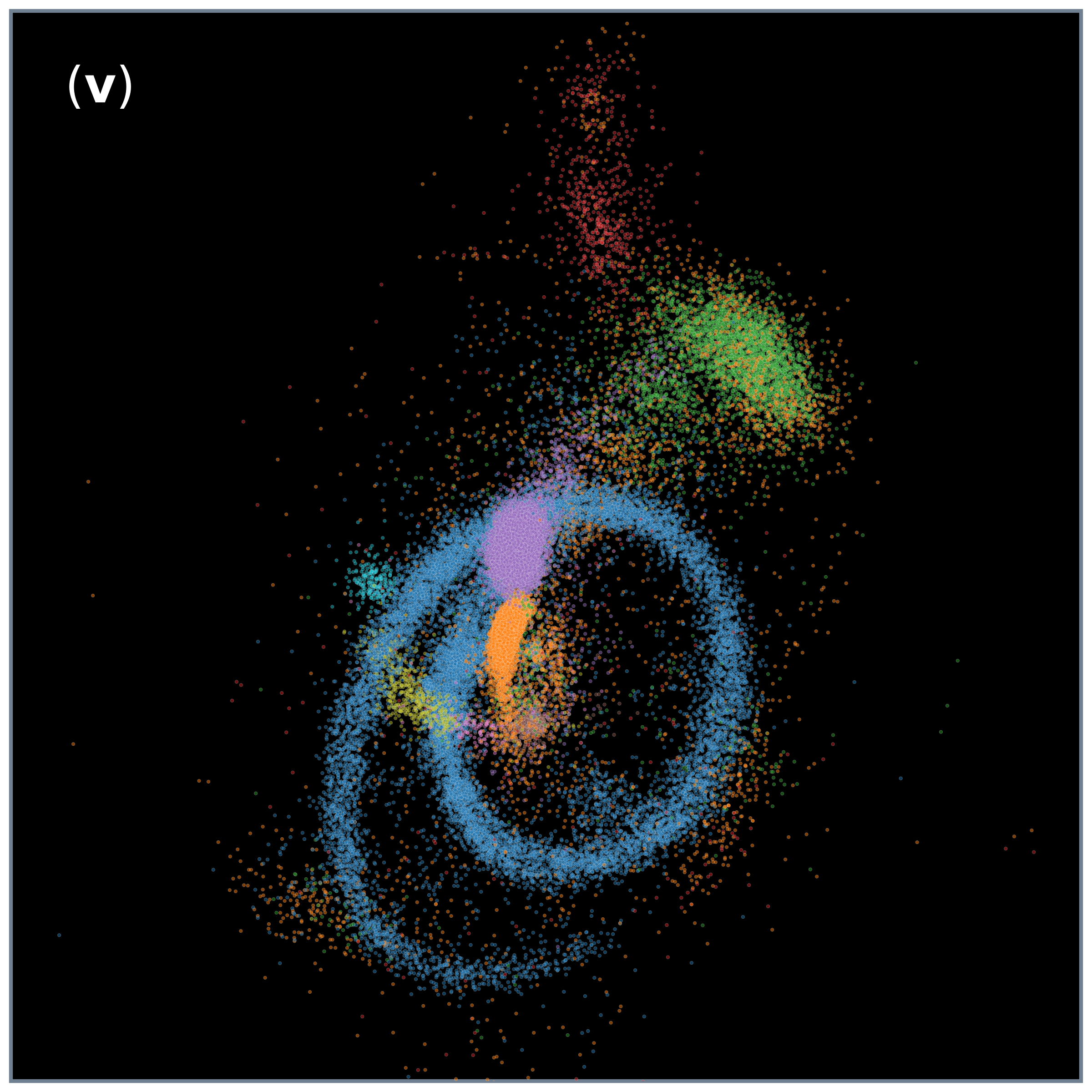}};
  %\pause
  \node (img3) at (img2.south west) [xshift = -1.25cm, yshift = -1.3cm] {\includegraphics[height=4.7cm]{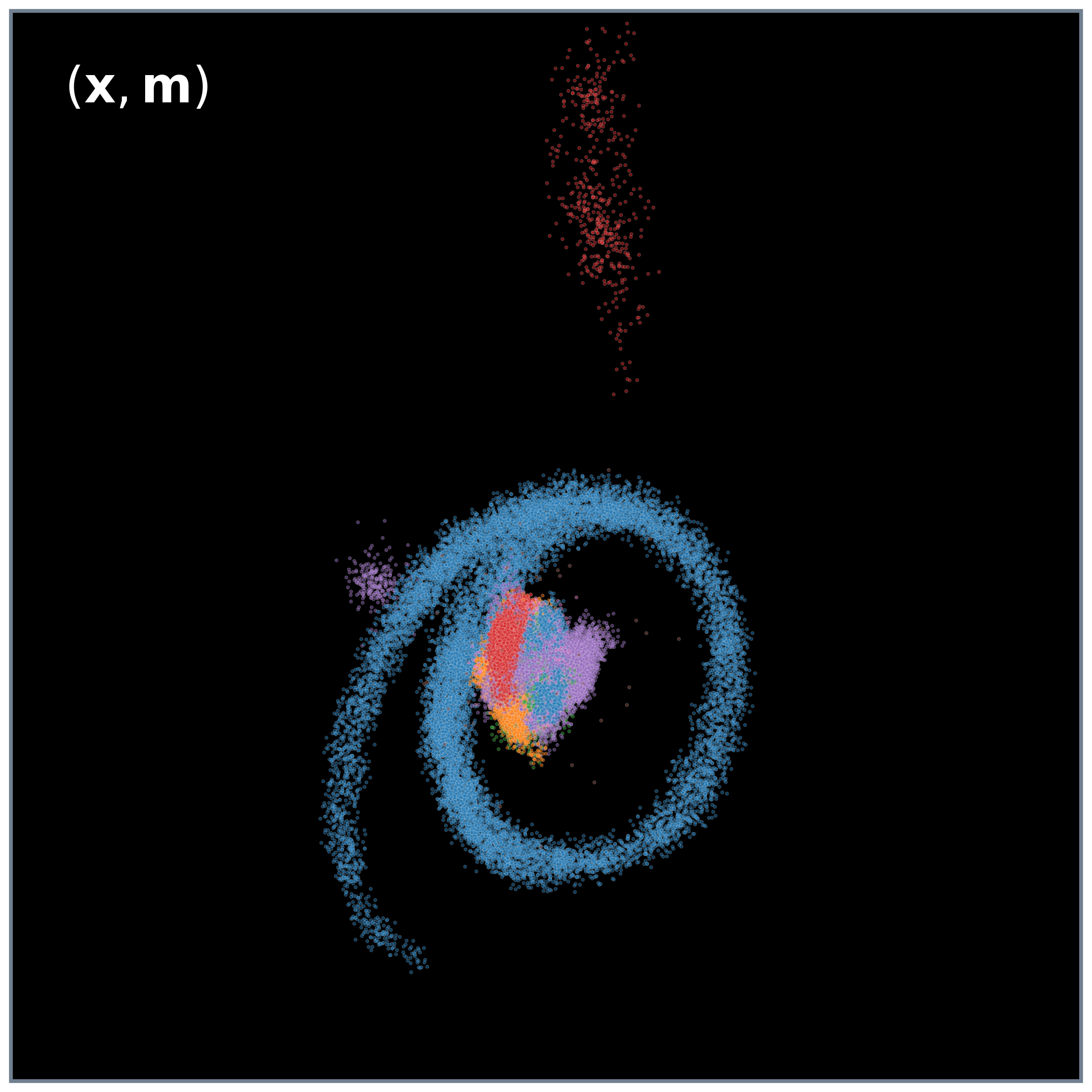}};
  %\pause
  \node (img4) at (img3.south east) [xshift = 1.25cm, yshift = 1.3cm] {\includegraphics[height=4.7cm]{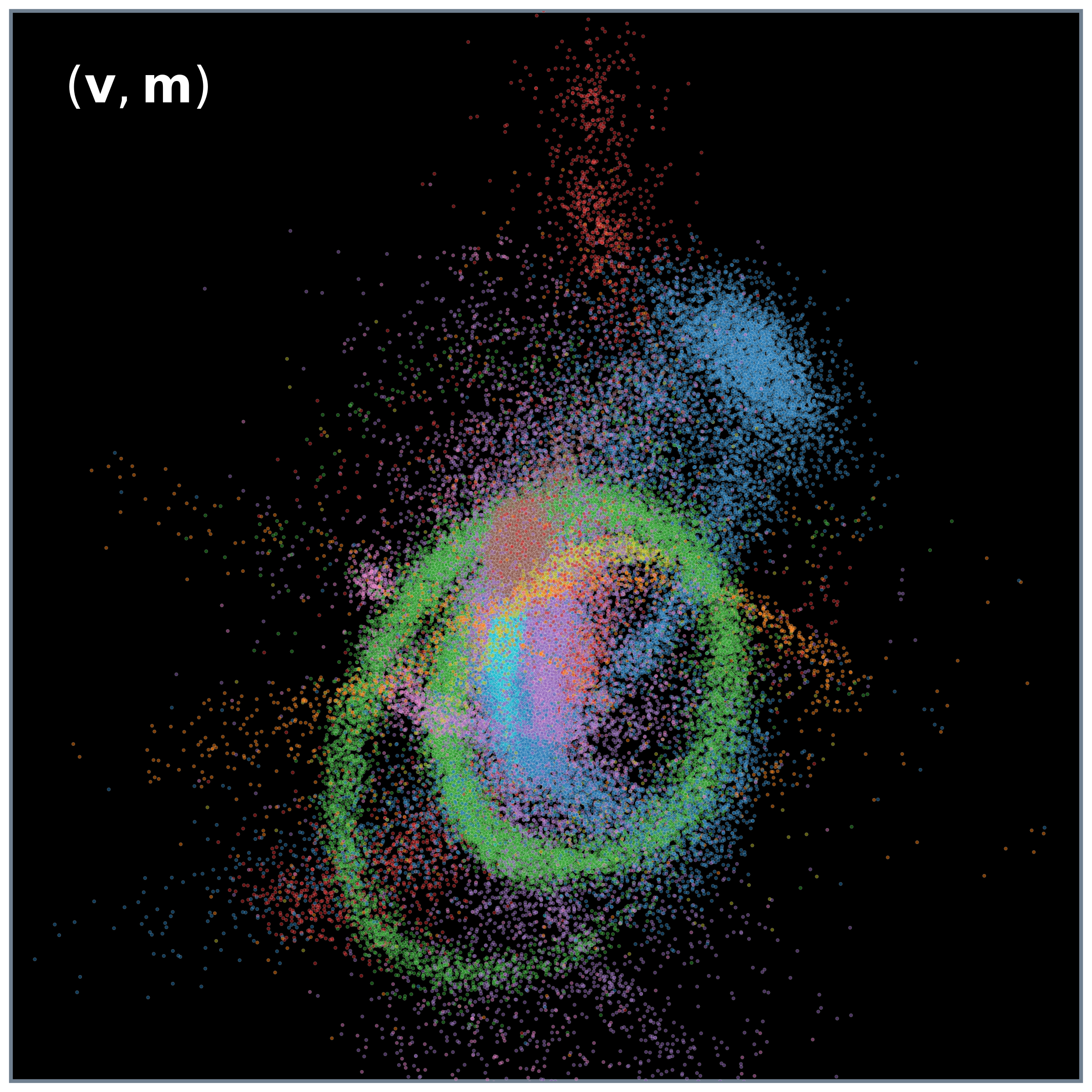}};
  %\pause
  \node (img5) at (img4.south west) [xshift = -1.25cm, yshift = -1.3cm] {\includegraphics[height=4.7cm]{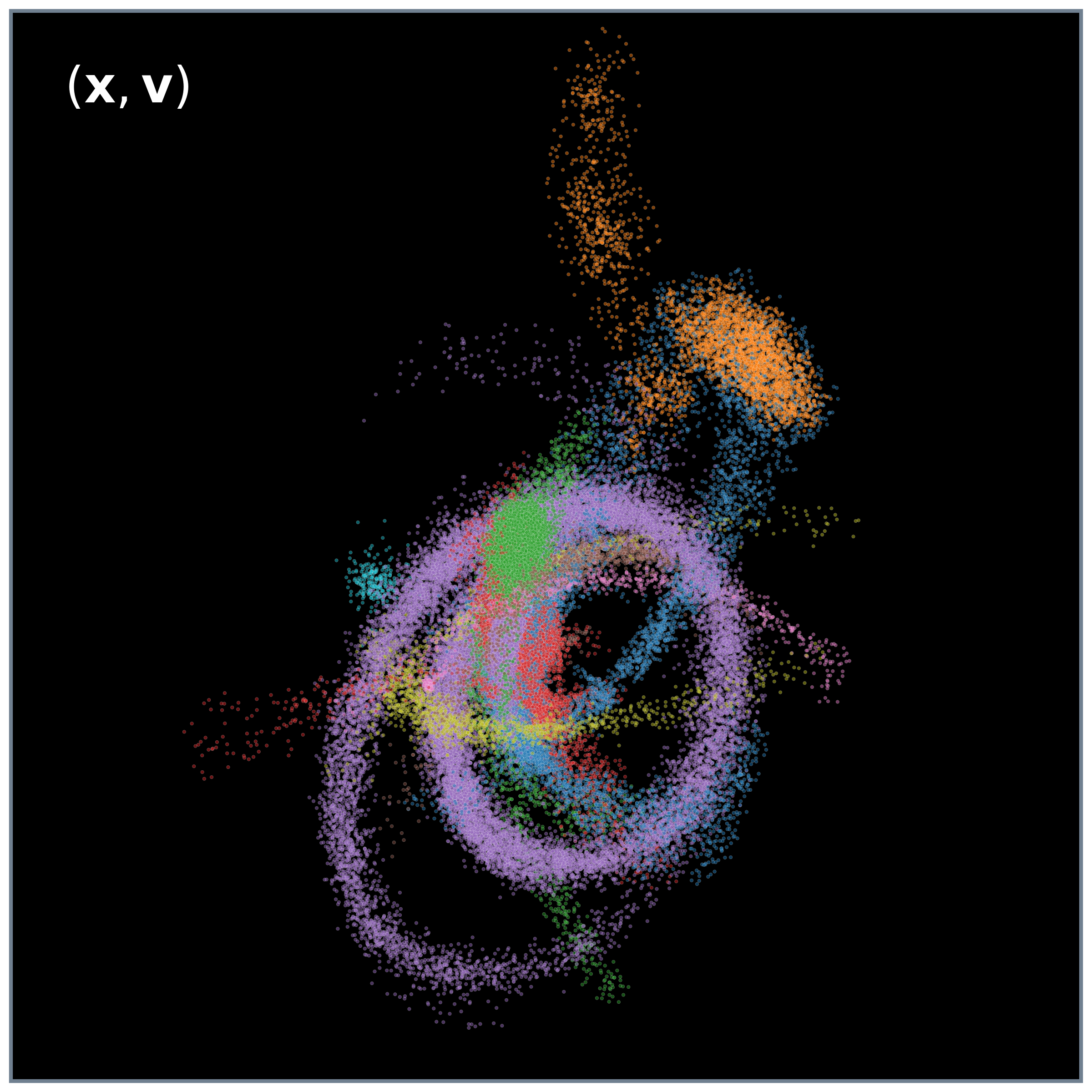}};
  %\pause
  \node (img6) at (img5.south east) [xshift = 1.25cm, yshift = 1.3cm] {\includegraphics[height=4.7cm]{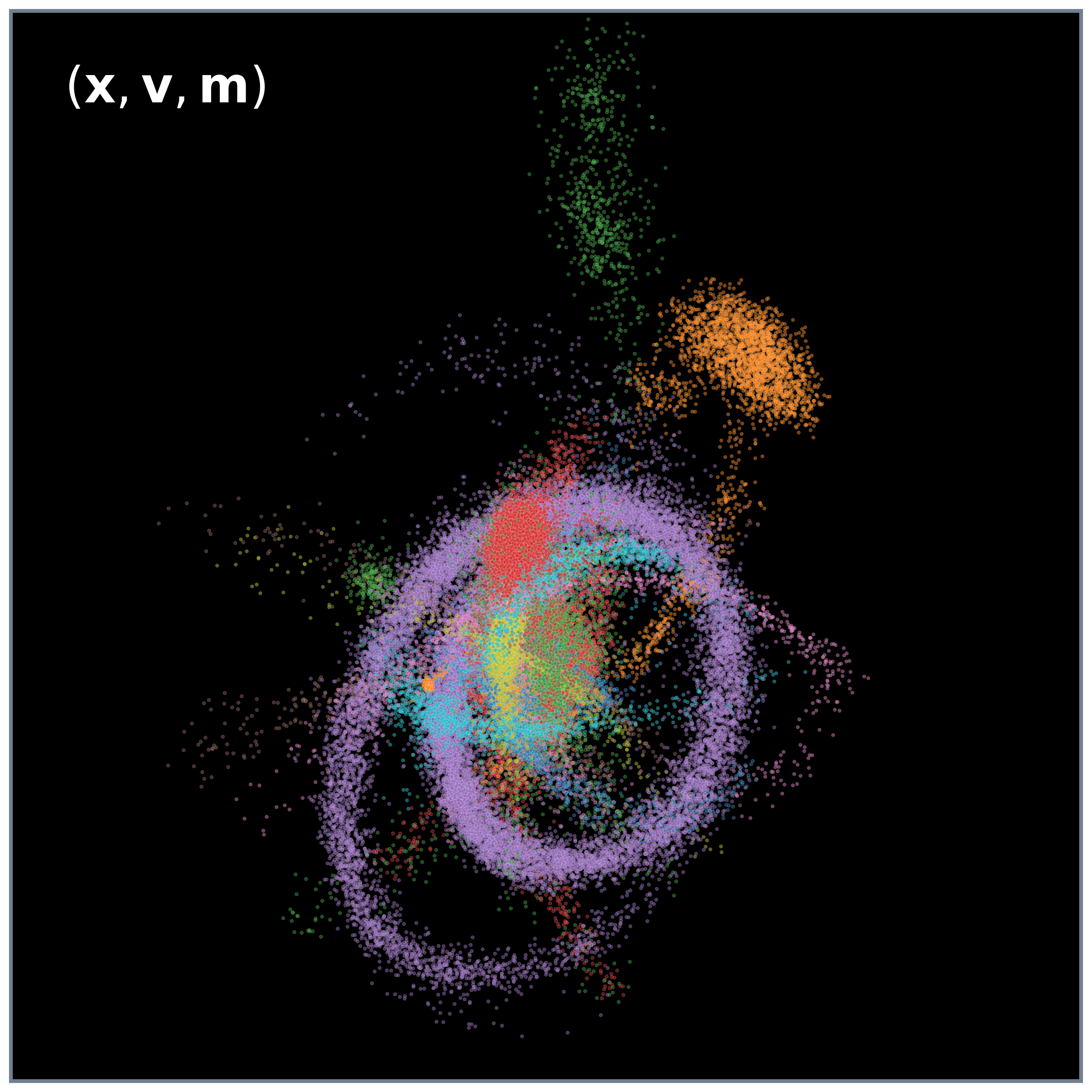}};
\end{tikzpicture}
\end{center}
\vspace{-0.4cm}
\caption{Illustrations of the spatial positions of clusters found by applying \astrolink\ to the artificial galaxy, newhalo\_young, over various feature spaces. Clusters found by \astrolink\ are shown with the same colours as in the ordered-density plots of Fig. \ref{fig:newhalo_young_ordered_density}.}
\label{fig:newhalo_young_3d_visuals}
\end{figure}

Fig. \ref{fig:newhalo_young_3d_visuals} depicts the spatial distributions of those clusters shown coloured in the corresponding ordered density plots of Fig. \ref{fig:newhalo_young_ordered_density}. Each panel gives an indication as to which types of clusters \astrolink\ can retrieve given that the input data is expressed through certain feature space combinations. We see here that spatially compact clusters are best described by spatial features, stream-like clusters are best described by kinematic and chemical features, and combining these feature sets tends to give good results for various cluster types -- permitting their retrieval simultaneously. Notably, the addition of chemical abundances to the input data's feature space gives \astrolink\ more knowledge about otherwise phase-mixed clusters. These observations agree with theoretical and intuitive predictions about how one would expect a generalised astrophysical clustering algorithm to perform.

\section{Conclusions} \label{sec:conclusion}
We have presented \astrolink, a generalised astrophysical clustering algorithm that produces a set of arbitrarily-shaped hierarchical clusters from an arbitrarily-shaped point-based data set such that the clusters found are statistical outliers from noisy density fluctuations. The algorithm is unsupervised, requires little to no input from the user (other than the trivial requirement to supply the data), and has an intuitive set of hyperparameters that can be quickly and intuitively adjusted to ensure that the resulting clusters meet the user's needs.

We have also demonstrated the robustness of \astrolink's performance in comparison to its predecessors, \clustar\ and \hoptics. We find that \astrolink\ classifies satellites from simulated galaxies with greater accuracy while requiring less run-time and memory allocation. Similarly to \hoptics, it also produces a visual representation of the input data's implicit clustering structure. However unlike \hoptics, the ordered-density plot is determinable and more easily interpretable -- owing to the fact that absolute vertical differences correspond directly to overdensity factors.

While there are still ways to improve \astrolink's output (e.g. defining a locally-adaptive metric before estimating density, propagating data point uncertainties into cluster uncertainties, using an supplementary model-based approach that attributes additional data points to the \astrolink\ clusters, etc.). \astrolink\ is well-suited to the clustering problem of finding astrophysically relevant groups from large-scale data sets in both simulated and observational settings.

\section*{Acknowledgements}
WHO acknowledges financial support from the Carl-Zeiss-Stiftung and the Paulette Isabel Jones PhD Completion Scholarship at the University of Sydney. TB's contribution to this project was made possible by funding from the Carl-Zeiss-Stiftung.

%%%%%%%%%%%%%%%%%%%%%%%%%%%%%%%%%%%%%%%%%%%%%%%%%%

\section*{Data Availability}
The data underlying this article may be made available on reasonable request to the corresponding author. \astrolink\ is publicly available at \url{https://github.com/william-h-oliver/astrolink}.


\bibliographystyle{mnras}
\bibliography{references}

%%%%%%%%%%%%%%%%%%%%%%%%%%%%%%%%%%%%%%%%%%%%%%%%%%

%%%%%%%%%%%%%%%%% APPENDICES %%%%%%%%%%%%%%%%%%%%%

\appendix

%%%%%%%%%%%%%%%%%%%%%%%%%%%%%%%%%%%%%%%%%%%%%%%%%%

% Don't change these lines
\bsp	% typesetting comment
\label{lastpage}
\end{document}